\documentstyle[epsf,A4]{article}

\include{donnees}

\begin{document}

\begin{center}

{\Large \bf FRONTS AND INTERFACES}

\vspace{0.1 cm}

{\Large \bf IN}

\vspace{0.1 cm}

{\Large \bf BISTABLE EXTENDED MAPPINGS}  

\vspace{0.4 cm}
{\bf Ricardo COUTINHO$^1$ and Bastien FERNANDEZ$^2$}

\vspace{0.4 cm}
$^1$ Departamento de Matem\'atica, Instituto Superior
T\'ecnico 

Av.\ Rovisco Pais 1096, Lisboa Codex Portugal
 
e-mail: rcoutin@math.ist.utl.pt

\vspace{0.4 cm}
$^2$ The Nonlinear Centre, Department of Applied
Mathematics and Theoretical Physics 

Silver Street, Cambridge CB3 9EW U.K.\ 

e-mail: b.fernandez@damtp.cam.ac.uk

\vspace{0.8 cm}
{\bf Abstract}
\end{center}
We study the interfaces' time evolution 
in one-dimensional bistable extended dynamical systems with discrete
time. The dynamics is governed 
by the competition between a local piece-wise affine bistable mapping
and any couplings given by the convolution with a function of bounded
variation. We prove the
existence of travelling wave interfaces, namely fronts, and the
uniqueness of the corresponding selected velocity and shape. This selected
velocity is shown to be the propagating velocity for any interface,
to depend continuously on the couplings and to increase with the
symmetry parameter of the local nonlinearity. We apply the results
to several examples including discrete and continuous
couplings, and the planar fronts' dynamics in
multi-dimensional Coupled Map Lattices. We eventually emphasize on the
extension to other kinds of fronts and to 
a more general class of bistable extended mappings for which
the couplings are allowed to be nonlinear and the local map to be smooth.

\vspace{0.1 cm}


\section{Introduction}
The fronts in bistable discrete dynamical systems defined on a
one-dimensional lattice have been investigated in various models such as
in Lattice Dynamical Systems (LDS) \cite{Afrai2}, or in Coupled Map Lattices
(CML) \cite{Carretero,Coutinho96}. These models can be viewed as
being in the class of (extended) dynamical systems for which the time evolution
is given by 
\begin{equation}\label{GEND}
u^{t+1}={\cal L}u^t+{\cal L}'F\circ u^t,
\end{equation}
where $u^t$ is a lattice configuration at time $t\in \Bbb N$ (or
$\Bbb Z$), and ${\cal L}$ and
${\cal L}'$ are some linear, continuous and homogeneous operators, the
so-called couplings. The mapping $F$ is the direct product of a
(local) bistable one-dimensional map $f$, i.e.\ the dynamics induced
by $f$ has two stable fixed points. In LDS ${\cal L}$ is often chosen
to be the Laplacian to model a diffusive process and ${\cal
L}'$ is a multiple of the identity \cite{Afrai1,Afrai2,Oppo}. On the
opposite, in CML the
diffusion is included in ${\cal L}'$ and ${\cal L}$ is generally
assigned to zero \cite{Carretero,Coutinho96,Chaos,Kaneko}. In all these
models, the dynamics is governed by the competition between the fixed
points along the lattice and then results in pattern formation.

To describe the fronts in these space-time discrete 
systems, one has to consider bounded configurations depending on a
spatio-temporal continuous variable, the
so-called (front) shapes \cite{Coutinho96}. It is then
useful to extend the domain of ${\cal L}$ and of ${\cal L}'$, and therefore the
phase space of these extended systems, to bounded functions of a
real variable. By doing so, one obtains a more general class of
extended dynamical systems with discrete time for which, to each point
of the real line is associated a bistable map, and these maps interact
through ${\cal L}$ and ${\cal L}'$. This new class contains in
particular LDS and CML. 

In this work, to obtain explicit expressions, the local
map is chosen to be piece-wise affine. However we shall see how the
existence of fronts can be stated for some bistable models with
a smooth, say $C^{\infty}$, local map. Up to a linear change of
variable, it is always possible to assume the two local fixed points
to be $0$ and $1$. 
The resulting expression is
then (see Figure 1)
\begin{figure}
\epsfxsize=6truecm 
\centerline{\epsfbox{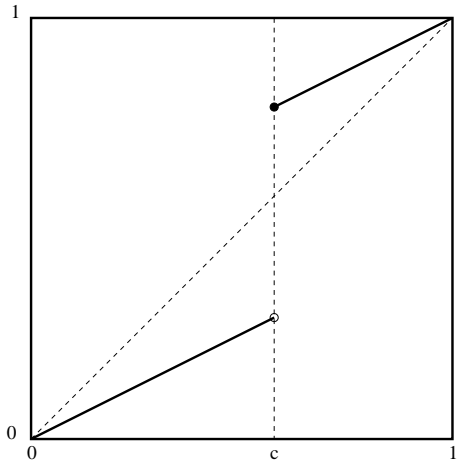}}
\caption{The local map $f$.}
\end{figure}
\begin{equation}\label{LOC}
f(x)=ax+(1-a)H_c(x),
\end{equation}
where $a\in [0,1)$, $c\in (0,1]$ and 
$$
H_c(x)=\left\{\begin{array}{l}
0\quad {\rm if}\quad x<c\\
1\quad {\rm if}\quad x\geq c
\end{array}\right. ,
$$
is the Heaviside function at $c$.\footnote{In all the paper, $H_{\omega}$
stands for the Heaviside function at $\omega$.}

By combining relations (\ref{GEND}) and (\ref{LOC}), one obtains the
general expression for a piece-wise affine bistable extended
mapping
\begin{equation}\label{DEFD}
u^{t+1}=L_1u^t+L_2H_c\circ u^t,
\end{equation}
The goal of this article is to describe the dynamics of fronts and of
interfaces in this
system. Some general conditions are imposed (below) to $L_1$ and to
$L_2$ in order to ensure 
the model to be interpreted as a bistable mapping with homogeneous couplings. 
We shall also see how the results
apply to the description of planar fronts in multi-dimensional CML. 

The plan of the article is as follows. We begin by defining the
dynamical system and particularly the couplings under
consideration. Still in Section 2, a representation of couplings in
terms of convolutions is given and several related properties are
reminded or established. The definition and the existence of fronts are
exposed in Section \ref{MAINSE}. 
In this section, we also investigate the set of
possible fronts velocities and the set of the symmetry parameter $c$
for the (non-)existence of fronts. 
Furthermore by introducing a distance in the set of functions
associated with the couplings, the selected velocity is shown to vary
continuously with the latter. In the following section, we prove that
any interface, i.e.\ any orbit for which the state at each
time is a configuration linking 
two different phases, always propagates with
the fronts velocity. Section 5 deals with applying 
this formalism to the analysis of the fronts velocity 
in a LDS and in the continuous
purely diffusive model. Moreover we describe how the planar fronts' dynamics
in multi-dimensional CML enters in this framework. Finally the
results stated in the plateaus, i.e.\ the intervals in $c$ for which
the velocity is constant, are extended to bistable extended
mappings with nonlinear couplings and smooth local maps. We also
prove the existence of other fronts as the unstable ones. 

\section{Definitions and preliminary results}
\subsection{The dynamical system}
First of all, the phase space is chosen to be the space, called ${\cal M}$,
of bounded Borel measurable functions on $\Bbb R$,
endowed with the uniform norm
$$
\|u\|=\sup_{x\in \Bbb R} |u(x)|. 
$$
In this phase space, the action of the dynamics on the orbits
$\{u^t\}_{t\geq 0}$ is given by the relation (\ref{DEFD}) where $L_1$
and $L_2$ are linear continuous operators satisfying the following
conditions 

\noindent
{\sl (i)} $L_1$ and $L_2$ are s-homogeneous 

\noindent
{\sl (ii)} $L_1$ and $L_2$ are positive 

\noindent
{\sl (iii)} $(L_1+L_2)\1=\1 $ and $L_2\neq 0$,

\noindent
where $\1$ stands for the function $u(x)=1$. For $\omega\in \Bbb R$ and $u\in
{\cal M}$, let
$$
\left(\sigma^{\omega}u\right)(x)= u(x-\omega)\quad \forall x\in\Bbb R ,
$$
be the shift (or translation) operator. An operator $L$ with domain ${\cal M}$
is said to be homogeneous if
\begin{equation}\label{HOM}
L\sigma^{\omega}=\sigma^{\omega}L\quad \forall \omega
\in \Bbb R ,
\end{equation}
and is said to be s-homogeneous if it is homogeneous and commutes with
the point-wise limit of equi-bounded sequences in ${\cal M}$, i.e.\ if
$\{u_n\}_{n\in\Bbb N}$, $u_n\in {\cal M}$ is such that
$$
\exists m\in \Bbb R\ :\ \sup_{n\in \Bbb N}\left\| u_n\right\| <m\quad 
{\rm and}\quad \forall x\in \Bbb R\ 
\lim_{n\rightarrow +\infty}u_n(x)=u(x),
$$
then
\begin{equation}\label{SHOM}
\forall x\in \Bbb R\ \lim_{n\rightarrow +\infty}\left(Lu_n\right)(x)
=\left(Lu\right)(x).
\end{equation}
The s-homogeneity reflects the homogeneous and the punctual
characteristics of the dynamics (\ref{DEFD}). An operator $L$ is positive
if $\forall u\in {\cal M}$ such that $u\geq 0$, we have $Lu\geq
0$.\footnote{$u\geq 0$ means $u(x)\geq 0$ $\forall x\in\Bbb R$.} The
condition {\sl (ii)}, together with {\sl (iii)}, indicates that the
dynamics (\ref{DEFD}) can be interpreted as the competition between a
local reaction and a global diffusive process. The condition {\sl
(iii)} guarantees the invariance of the local fixed points by the
couplings.

After having defined the dynamics
using s-homogeneous operators, we now obtain an integral
representation of the latter using the Lebesgue-Stieltjes
integral. The following statement reveals the existence of a bijection
between the s-homogeneous operators on ${\cal M}$ and the right
continuous functions of bounded variation vanishing at
$-\infty $. This set of functions will be denoted by $BV_{0}$.
\begin{Pro}\label{REPR}
A linear and continuous operator $L$ of ${\cal M}$ into itself is 
s-homogeneous iff it exists a unique function $h\in BV_0$, such that
\begin{equation}\label{INTE}
\forall u\in {\cal M},\quad 
(Lu)(x)=\int_{\Bbb R} u(x-y)dh(y) \quad
\forall x\in \Bbb R.
\end{equation}
Moreover $L$ is positive iff $h$ is increasing.
\end{Pro}

\noindent
{\sl Proof:} 
If the integral $(\ref{INTE})$ defines $L$, then the Lebesgue dominated
convergence theorem shows that $L$ is s-homogeneous.
Conversely let $h(x) =LH_0(x)$. Then by relations 
$(\ref{HOM})$ and $(\ref{SHOM})$, 
$h$ is right continuous and ${\displaystyle \lim_{x\rightarrow 
-\infty}}h(x)=0$. By using
$(\ref{HOM})$ and the linear continuity of $L$, we conclude
that $h$ is of bounded variation in $\Bbb R$ (similarly as in the
proof of the Riesz theorem in \cite{Kolmo}). 
Then $(\ref{INTE})$ is 
true for all the functions of the form
$$
u=\sum_{n=1}^Ns_n\left(H_{x_{n-1}}-H_{x_{n}}\right).
$$
Any continuous function is the point-wise limit of a sequence of functions of
this form. Relation $(\ref{INTE})$ then holds in the (Baire) 
class ${\cal B}^0$ 
of all bounded continuous functions. By the same argument, it is also
valid in the class ${\cal B}^1$ of functions which are 
point-wise limits of functions in ${\cal B}^0$. We can proceed
in this way transfinitely to conclude that the
proposition is true for any bounded function in a Baire class 
${\cal B}^{\alpha }$ (where $\alpha $ is a countable ordinal). But 
$\bigcup\limits_{\alpha }{\cal B}^{\alpha }$ is exactly the set of all Borel
measurable functions \cite{Poussin}.\footnote{Requiring the functions
to be bounded is not restrictive because any unbounded Borel measurable 
function is a point-wise limit of bounded Borel measurable
functions. Hence we can consider the classes ${\cal B}^{\alpha }$ containing
only bounded functions.} \hfill $\Box $

Let us recall that, for a function $h\in BV_0$, 
the convolution is defined by 
$$
\forall u\in {\cal M} \quad h\ast u(x)=\int_{\Bbb
R}u(x-y)dh(y)\quad \forall x\in \Bbb R .
$$ 
We have adopted an unusual ordering in this definition to indicate that
the convolution by $h\in BV_0$ corresponds to the
action of an operator in ${\cal M}$. Moreover notice that the convolution
$h\ast u$ is defined even if $u\not\in BV_0$ and is thus in general
non-commutative. 

We then obtain an equivalent expression for the dynamical system (\ref{DEFD})
\begin{equation}\label{DEFD2}
u^{t+1}=h_1\ast u^t+h_2\ast H_c\circ u^t,
\end{equation}
where $h_1$ (resp.\ $h_2$) corresponds to $L_1$ (resp.\ $L_2$) 
according to Proposition \ref{REPR} and is thus an increasing
function in $BV_0$. We denote by
${\cal I}$ the subset of $BV_0$ and of ${\cal M}$ composed with right 
continuous increasing functions vanishing at $-\infty$. This
alternative expression for the dynamical system, combined with the
properties of the convolution and of the functions in ${\cal I}$,
both described in the following section, will prove to be convenient for
the analysis of fronts. To conclude this section, we 
mention that the interfaces' dynamics in a 
similar but space-time continuous bistable extended dynamical
system, i.e.\ a first order in time differential equation for which
the interaction is given by the convolution with a $C^1$ function, has
been investigated in \cite{Bates}.

\subsection{Basic properties}
Using Fubini theorem, it is shown that the convolution is commutative
and associative in $BV_0$. As a consequence,
two any linear continuous and s-homogeneous operators commute. 

The following properties are a direct consequence of the definition

\noindent
{\sl (i)} If $h_1,h_2\in {\cal I}$, then $h_1\ast h_2\in {\cal I}$ and
$\|h_1\ast h_2\|=\|h_1\|\|h_2\|$.

\noindent
{\sl (ii)} Let $h\in {\cal I}$ and 
$u_1,u_2\in {\cal M}$. If $u_1\leq u_2$,\footnote{i.e.\ if $u_1(x)\leq u_2(x)$ 
$\forall x\in \Bbb R$} then $h\ast u_1\leq h\ast u_2$.

\noindent
{\sl (iii)} $\forall u\in {\cal M} \quad \sigma^{\omega}u=H_{\omega}\ast u$,

\noindent
{\sl (iv)} $\forall \omega,\omega'\in \Bbb R\quad 
H_{\omega}\ast H_{\omega'}=H_{\omega+\omega'}$.

Moreover let the left closure of a real subset $A$ be defined
by\footnote{ $A+B=\left\{ x+y\ :\ x\in A\ {\rm and}\ y\in B \right\}$}
$$
{\lclos A}=\bigcap_{\delta >0} A+[0,\delta).
$$
Using the diagonal process, one shows two basic properties of the left
closure.

\noindent
{\sl (v)} ${\lclos{{\lclos A}+{\lclos B}}}={\lclos{A+B}}$.

\noindent
{\sl (vi)} ${\lclos{{\displaystyle\bigcup_{i\in \Bbb N}}{\lclos{A_i}}}}= 
{\lclos{{\displaystyle\bigcup_{i\in \Bbb N}}A_i}}$.

We now introduce two sets
which (partly) characterize the right continuous increasing
functions. For the interpretation in terms of coupling, they
determine the sets of points coupled to 0 and thus, using the
homogeneity, the generalized lattice on which the coupling acts.
For $h\in {\cal I}$ the set of increase points is defined
as follows.
$$
E(h)=\left\{ x\in \Bbb R\ : \ \forall \delta >0 \quad h(x)>h(x-\delta)
\right\}.
$$
The set of discontinuity points, a subset of $E(h)$, is given by
$$
D(h)=\left\{x\in \Bbb R\ : \ h(x)>h(x-0)\right\},
$$
where $h(x-0)={\displaystyle \lim_{\delta\rightarrow 0, 
\delta >0}}h(x-\delta)$. As a first result, one has a sufficient
condition for $E(h)$ to reduce to $D(h$).
\begin{Lem}\label{LEDE}
If $h\in {\cal I}$ and $E(h)$ is countable then $E(h)=D(h)$.
\end{Lem}

\noindent
{\sl Proof:} If $h\in {\cal I}$, then it is equal to the sum of a continuous 
function $h_c$ and a step function $h_d$ which are such that
$h_c,h_d\in {\cal I}$ and $D(h)=E(h_d)=D(h_d)$ \cite{Kolmo}. 
Hence $E(h)\setminus D(h)\subset E(h_c)$. Now by assumption
let $\{a_n\}_{n\in\Bbb N}$ be an enumeration of $E(h_c)$. We obtain 
$$
\int_{E(h_c)}dh_{c}=\sum_{n\in\Bbb N}h_c(a_n)-h_c(a_n-0)=0,
$$
because $\mu_h(E(h)^c)=0$, where $\mu_h$ is the Lebesgue-Stieltjes
measure associated with $h\in {\cal I}$, and where $A^c$
denotes the complement of $A$ in $\Bbb R$. 
As a consequence $h_c=0$ and $h=h_d$. \hfill $\Box $

Moreover, the evolution of these two sets under the convolution and
the summation is given in the following statement.
\begin{Pro}\label{CONV}
{\rm (1)} Let $h,h'\in {\cal I}$. Then $h\ast h'\in {\cal I}$ and 
$$
\begin{array}{ccl}
{\rm (1a)} \quad E(h\ast h')&=&{\lclos{E(h)+E(h')}},\\
{\rm (1b)} \quad D(h\ast h')&=&D(h)+D(h').
\end{array}
$$ 

\noindent
{\rm (2)} Let $\{h_i\}_{i\in\Bbb N}$ be a uniformly summable sequence of
functions in ${\cal I}$. Then $h={\displaystyle
\sum_{i\in \Bbb N}}h_i\in {\cal I}$ and 
$$
\begin{array}{ccl}
{\rm (2a)} \quad E(h)&=&{\lclos{{\displaystyle \bigcup_{i\in\Bbb N}}E(h_i)}},\\
{\rm (2b)} \quad D(h)&=&{\displaystyle \bigcup_{i\in\Bbb N}}D(h_i).
\end{array}
$$ 
\end{Pro}

\noindent
{\sl Proof:} We only show relation (1a) here. Relation (1b)
can be proved similarly. Statement (2b) is a direct consequence
of uniform convergence and (2a) follows from (\ref{CROIS}) below. 

\noindent
For an increasing function, the definition of
$E(h)$ implies that for a given $\delta>0$,
\begin{equation}\label{CROIS}
h(x)-h(x-\delta)> 0\quad {\rm iff}\quad (x-\delta,x]\cap E(h)\neq
\emptyset.
\end{equation}
Consequently given $h,h'\in {\cal I}$ we have for any 
$x\in \Bbb R$ and $\delta >0$
$$
h\ast h'(x)-h\ast h'(x-\delta)=
\int_{\left(-E(h)+(x-\delta,x]\right)\cap E(h')}
h(x-y)-h(x-y-\delta)dh'(y),
$$
from which we deduce the relation
$$
\forall \delta>0\quad h\ast h'(x)-h\ast h'(x-\delta)>0\quad {\rm
iff}\quad \forall \delta>0\ x\in E(h)+E(h')+[0,\delta).
$$
This equivalence is nothing but (1a). \hfill $\Box$

\section{The fronts}\label{MAINSE}
With the previous definitions and properties provided, we are now able to
analyse the dynamics of the simplest structures linking two 
different phases\footnote{The phases of the system are the fixed
points 0 and 1 which compete together when coupled.}, namely the
fronts. In this section the latter 
are firstly defined. Then we state their existence, in
particular by considering various conditions on the couplings $h_1$
and $h_2$ and on the parameter $c$. We also prove the continuous
dependence of their velocity on the couplings using an adequate topology. We
finally consider the existence of anti-fronts by relating it to the
fronts' existence.

\subsection{The existence of fronts}
\begin{Def}
A front of velocity $v$ is an orbit of (\ref{DEFD2}) defined by
$$
u^t=H_{vt+x_0}\ast\phi\quad \forall t\in\Bbb Z,
$$
where $x_0\in\Bbb R$ and the shape $\phi\in {\cal M}$ is such that
\begin{equation}\label{EXIST}
H_c\circ \phi=H_o.
\end{equation}
\end{Def}
This definition is somehow restrictive because, for the sake of
simplicity in this section, it imposes $\phi(0)\geq c$. However, we
shall see in Section \ref{OMODEL} how to relax this condition by imposing the
shape to satisfy, instead of (\ref{EXIST}), the condition $\phi(x)<c$ if $x<0$
and $\phi(x)\geq c$ if $x>0$, and therefore by letting $\phi(0)$
arbitrary.

Because of the translational invariance in (\ref{DEFD2}), if it exists a front,
then there is also an infinite number of fronts with the same velocity
and the same shape, one for each value of $x_0$.

As we shall see below in the proof of Theorem \ref{MAINR}, 
this definition implies that the shape $\phi$ is a right continuous
increasing function which depends on the velocity. Moreover it
behaves asymptotically as 
$$
\lim_{z\rightarrow -\infty}\phi(z)=0\quad {\rm and}\quad 
\lim_{z\rightarrow +\infty}\phi(z)=1.
$$
The fronts thus defined are actually the travelling waves
resulting from the motion of an interface between the two different
phases of the dynamical system under consideration. 

The natural questions about the fronts deal with their existence, their
uniqueness and the selected velocity
given two any couplings $h_1$ and $h_2$ and a value of the symmetry
parameter $c$. It is
also of interest to determine the set $P$ of possible velocities
when $c$ varies in $(0,1]$, and the couplings are kept
fixed. 
\begin{Thm}\label{MAINR}
Given two couplings $h_1$ and $h_2$, there is a Baire first category set $G$ 
such that for any $c\in (0,1)\setminus G$, the dynamical system
(\ref{DEFD2}) has a unique front shape. The corresponding fronts have a
velocity $\bar{v}(c)$ which increases with $c$ and which can be
extended to a left continuous function on $(0,1)$. The set $P$ of possible
velocities is a dense subset of the
range of the function $\bar{v}$ and has the following bounds
$$
\inf P=v_{\rm min}\quad {\rm and}\quad \sup P=v_{\rm max},
$$
where $v_{\rm min}=\min\left\{\inf E(h_1),\inf E(h_2)\right\}$ and 
$v_{\rm max}=\max\left\{\sup E(h_1),\sup E(h_2)\right\}$. 
\end{Thm}
Assuming some additional (general) conditions on $h_1$ and $h_2$, the
set $G$ is shown to be countable and nowhere dense (see section 
\ref{SECTG}). If $v_{\rm min}=\inf E(h_1)$
and $v_{\rm max}=\sup E(h_1)$, then $P$ is dense in $(v_{\rm
min},v_{\rm max})$ and therefore $\bar{v}$ is a continuous function of
$c$ (see Section \ref{SECTP}). If the function $h_2$ is 
continuous, then $G$ is empty, $P$ contains $(v_{\rm min},v_{\rm
max})$, $\bar{v}$ is continuous and the shape $\phi$ also varies
continuously with its argument. 

For $c\in G$ no front exists. We have instead a ``ghost front'', i.e.\
a sequence of configurations in ${\cal I}$ which is not an orbit of 
(\ref{DEFD2}) but which attracts a set of initial conditions
\cite{Coutinho96}.
Ghost fronts are due to the discontinuous character of the
local map possibly combined with the couplings. 
The ghost front's velocity $\bar{v}(c)$ may belong to $P$ but it
may not since, in general, $P$ and the range of $\bar{v}(.)$ are different.
If $\bar{v}(c)$ belongs to $P$ this ghost orbit can be avoided by changing the
value of the local map at $c$ for fixed couplings.

\subsection{Proof of the fronts' existence}
The proof of Theorem \ref{MAINR} consists in investigating 
the function $\phi$, which can be defined independently
of the front existence, to deduce the conditions for it to be a front
shape.

Assuming the existence of a front of velocity $v$, we obtain from
(\ref{DEFD2})
$$
H_v\ast\phi=h_1\ast\phi+h_2\ast H_0.
$$
Then by using the convolution properties, we solve this equation to
obtain the following expression, where the dependence on $v$ has been added
\begin{equation}\label{FRON}
\phi(z,v)=\sum_{k=0}^{\infty}h_2\ast h_1^{\ast k}(z+(k+1)v)\quad 
\forall z\in\Bbb R,
\end{equation}
where for any $h\in BV_0$ we denote 
$$h^{\ast 0}=H_0\quad {\rm and}\quad 
h^{\ast (k+1)}=h^{\ast k}\ast h\quad {\rm for}\ k\geq 0.
$$
Consequently a front of velocity $v$ exists iff the function
defined by (\ref{FRON}) satisfies condition (\ref{EXIST}). From
relation (\ref{FRON}), the convolution properties,  the condition {\sl
(iii)} on the couplings and Proposition \ref{CONV}, it turns out that
\begin{equation}\label{INCR}
\forall v\in\Bbb R \quad \phi(.,v)\in {\cal I},
\end{equation}
and 
$$
E(\phi(.,v))={\lclos{\bigcup_{k=0}^{\infty}E(h_2)++_kE(h_1)+\{-(k+1)v\}}},
$$
where for any real subset $A$ we use the notation 
$$+_0A=\{0\}\quad {\rm and}\quad +_{(k+1)}A=+_kA+A\quad {\rm for}\ k\geq 0.
$$
Similarly we have 
\begin{equation}\label{INCV}
\forall z\in\Bbb R \quad \phi(z,.)\in {\cal I},
\end{equation}
and 
$$
E(\phi(0,.))={\lclos{\bigcup_{k=0}^{\infty}\frac{E(h_2)++_kE(h_1)}{k+1}}},
$$
where for any $y\in \Bbb R\setminus \{0\}$
$$
\frac{A}{y}=\{\frac{x}{y}\ :\ x\in A\}.
$$
Property (\ref{INCR}) gives the following necessary condition for a 
front of velocity $v$ to exist for some $c\in (0,1]$
\begin{equation}\label{ZINE}
0\in E(\phi(.,v)).
\end{equation}
Actually if $0\not\in E(\phi(.,v))$, then 
$$
\exists \delta>0\ :\ \forall \epsilon\in (0,\delta)\quad
\phi(-\epsilon,v)=\phi(0,v),
$$
and the condition (\ref{EXIST}) cannot be satisfied for any $c$.
Conversely by choosing $c=\phi(0,v)$, we conclude that for all $v$
for which relation (\ref{ZINE}) holds, there is (at least) 
a value of $c$ for which the fronts of velocity $v$ exist. 
The set of possible velocities is then given by 
$$
P=\left\{v\in\Bbb R\ :\ 0\in E(\phi(.,v))\right\}.
$$
This set has the following properties 
\begin{equation}\label{INCLU}
\bigcup_{k=0}^{\infty}\frac{E(h_2)++_kE(h_1)}{k+1}\subset 
P\subset E(\phi(0,.)),
\end{equation}
from which the bounds on $P$ are deduced.\footnote{Since
$E(h)={\lclos{E(h)}}$, we have $\sup E(h)\in E(h)$ if $\sup
E(h)\neq\infty$. Consequently  
$v_{\rm max}\in P$ iff $\sup E(h_1)\leq \sup E(h_2)<+\infty$ and in
all (finite) cases
$v=\sup E(h_1)+\frac{\sup E(h_2)-\sup E(h_1)}{k+1}\in P$ 
for any $k\geq 0$.}

Now one has to determine which fronts of possible velocity are
selected given a value of the parameter $c$. From relation
(\ref{FRON}) it turns out that 
if $v_1<v_2$ then $\phi(0,v_1)\leq \phi(-(v_2-v_1),v_2)$, and thus if
$c\leq \phi(0,v_1)$ then the existence condition (\ref{EXIST}) shows
that $v_2$ cannot be the fronts velocity for this value of
$c$. According to this property, the velocity of the (assumed) existing fronts
is uniquely given by\footnote{$\bar{v}(c)$ is defined at $c=1$ iff 
$v_{\rm max}<\infty$. Indeed, if $v_{\rm max}=\infty$, the set 
$\left\{ v\in\Bbb R\ : \ \phi(0,v)\geq 1\right\}$ is empty.}
\begin{equation}\label{BARV}
\bar{v}(c)=\min\left\{v\in \Bbb R\ : \ \phi(0,v)\geq c\right\}.
\end{equation}
The corresponding fronts exist iff 
$$
\forall \delta >0\quad \phi(-\delta,\bar{v}(c))<c,
$$
and the non-existence set is given by
$$
G=\left\{c\ :\ \exists \delta >0\quad c=\phi(-\delta,\bar{v}(c))\right\}.
$$

From its definition and from relation (\ref{INCV}) the function $\bar{v}$
is clearly increasing. To show that it is left continuous, we notice
that given $\delta>0$ one has $\phi(0,\bar{v}(c)-\delta)<c$. Let then $c'$ 
be such that $\phi(0,\bar{v}(c)-\delta)<c'<c$. Again from the
definition, we obtain $\bar{v}(c)-\delta<\bar{v}(c')$ and consequently
$\bar{v}$ is left continuous since it is increasing. Moreover its
range is $E(\phi(0,.))$ and thus relation (\ref{INCLU}) shows that $P$
is a dense subset of this range. 

To achieve the proof we have to show that $G$ is of Baire first
category. To that goal let 
$$
G^{(n)}=\left\{c\ :\ c=\phi(-\frac{1}{n}-0,\bar{v}(c))\right\}. 
$$
We have 
$$
G=\bigcup_{n=1}^{\infty }G^{(n)},
$$
and it then remains to show that each $G^{(n)}$ is nowhere dense. By
contradiction, we assume that $G^{(n)}$ is dense in the
(non-degenerated) interval $(c_1,c_2)$. Then $(c_1,c_2]\subset
G^{(n)}$ since both the functions
$\bar{v}$ and $\phi(-\frac{1}{n}-0,.)$ are left continuous and increasing.
It follows that $(c_1,c_2]\subset G$. Now for any 
$c\in(c_1,c_2)$ we have 
$$
\phi(0-0,\bar{v}(c_1))\leq c_1<c=\phi(0-0,\bar{v}(c))<c_2=
\phi(0-0,\bar{v}(c_2)),
$$ 
that is to say $\bar{v}(c)\in\left(\bar{v}(c_1),\bar{v}(c_2)\right)$
and $\bar{v}(c)\not\in P$. This is impossible since $\bar{v}$ is
increasing and $P$ is a dense
subset of the range of $\bar{v}$. \hfill $\Box $

\subsection{Characterization of the velocity set $P$}\label{SECTP}
We now proceed to the analysis of $P$ and to its consequences. 
In the following lemma we firstly describe
$P$ inside and outside the convex hull of $E(h_1)$. Let us define 
$$
i_j=\inf E(h_j)\quad {\rm and} \quad s_j=\sup E(h_j)\quad {\rm for}\ j=1,2.
$$
\begin{Lem}\label{POSS}
(i) If $\# E(h_1)>1$ then for any $\theta\in (0,1)$, $\theta$
irrational, we have
$$\forall a,b \in E(h_1),\quad a+\theta (b-a)\in P.$$

\noindent
(ii) If $s_1<s_2<+\infty$ then
$$
P\cap(s_1,s_2]=E(\phi(0,.))\cap (s_1,s_2].
$$

\noindent
(iii) If $\# E(h_1)>1$ and $s_2=+\infty$ then for any $\theta>1$,
$\theta$ irrational, we have 
$$\forall a,b \in E(h_1),\ a<b\ :\quad a+\theta (b-a)\in P.$$
\end{Lem}
The proof is given in Appendix \ref{FPROO}. Naturally one has statements
similar to {\em (ii)} when $-\infty<i_2<i_1$ and to {\em (iii)} when 
$i_2=-\infty$. Namely

\noindent
{\em (ii')} If $-\infty<i_2<i_1$ then
$$
P\cap[i_2,i_1)=E(\phi(0,.))\cap [i_2,i_1).
$$

\noindent
{\em (iii')} If $\# E(h_1)>1$ and $i_2=-\infty$ then for any $\theta>1$,
$\theta$ irrational, we have 
$$\forall a,b \in E(h_1),\ a<b\ : \quad b+\theta (a-b)\in P.$$

Moreover since $\bar{v}$ is a left continuous increasing function, a
discontinuity of $\bar{v}$ means the absence of an interval in its
range. As $P$ is contained in this range, statement {\em (i)} of the
previous lemma implies the following. 
\begin{Cor}
In the interval $\bar{v}^{-1}\left((i_1,s_1)\right)$, the selected 
velocity $\bar{v}(.)$ is a continuous function of $c$. 
\end{Cor}

Furthermore still in statement {\em (i)} of Lemma \ref{POSS}, we have
seen that all the irrational velocities in $(i_1,s_1)$ are 
fronts velocities. We now give a sufficient condition for all the 
rational velocities in this interval to be fronts velocities.
To that goal the lattice generated by a real subset $A$ is introduced
$$
R(A)=\overline{A+\bigcup_{k=0}^{\infty}+_k(A-A)},
$$
where $\overline{A}$ denotes the (usual) closure of $A$. Some
characterizations  
of the lattices generated by real subsets are listed in the following remark. 
\begin{Rem}\label{RESR}
(i) If $R(A)\neq \Bbb R$ and $\# A >1$, then it exists $\alpha\in\Bbb R^+$ 
such that $R(A)=a+\alpha\Bbb Z$ for any $a\in A$. 

\noindent
(ii) If $A$ contains an infinite bounded set, then $R(A)=\Bbb R$.

\noindent
(iii) If there are $x_0,x_1,x_2\in A$ such that $\frac{x_1-x_0}{x_2-x_0}
\in \Bbb R\setminus \Bbb Q$, then $R(A)=\Bbb R$.

\noindent
(iv) If $E(h)\neq D(h)$, then $R(E(h))=\Bbb R$ (see Lemma \ref{LEDE}).
\end{Rem} 
\begin{Pro}\label{ALLV}
If $E(h_2)\cap R(E(h_1))\neq \emptyset$, then $(i_1,s_1)\subset P$.
\end{Pro}
In particular, if one of the last conditions in Remark \ref{RESR} holds
for $E(h_1)$ then $(i_1,s_1)\subset P$ independently of $h_2$. 

\noindent
{\sl Proof:} Let $d\in E(h_2)\cap R(E(h_1))$ and $\delta >0$.
There
are $l\geq 1$, $\{a_n\}_{0\leq n\leq l}$, $a_n\in E(h_1)$,
$a_n<a_{n+1}$, and $\{m_i\}_{0\leq i<l}$, $m_i\in \Bbb Z$,
such that
\begin{equation}\label{D1}
d\leq a_0+\sum_{i=0}^{l-1}m_i(a_{i+1}-a_i)<d+\frac{\delta}{2},  
\end{equation}
We can suppose that $a_0$ is arbitrarily close to $i_1$ and $a_l=s_1$ 
by choosing, if necessary, $m_0=1$ and $m_{l-1}=0$. Now let
$J=2{\displaystyle \max_{0\leq i\leq l-1}}|m_i|$ and if $l>1$ let 
$M,N\in \Bbb Z^+$ be such that
\begin{equation}\label{D2}
0\leq N(a_l-a_{l-1})-M\sum_{i=0}^{l-2}(l-i)J(a_{i+1}-a_{i})<\frac{\delta}{2}.
\end{equation}
If $l=1$ we set $N=M=0$. 
Let us choose any $0<p<q$ co-prime integers and define
$$
n_{l-1}=-m_{l-1}-N+jp\quad {\rm and}\quad n_i=-m_i+M(l-i)J+jp\quad
{\rm for}\quad 0\leq i\leq l-2,
$$
where the integer $j$ is chosen large enough so that
$$
jq-1\geq n_0\geq n_1\geq \cdots \geq n_{l-1}\geq 0.
$$
By adding $(\ref{D1})$ and $(\ref{D2})$ we obtain
\begin{equation}\label{REZO}
d\leq a_0-\sum_{i=0}^{l-1}(n_i-(k+1)\frac{p}{q})(a_{i+1}-a_i)<d+\delta,  
\end{equation}
where $k=jq-1$. It is easy to see that the latter is equivalent to
$$
-\delta <d+(k-n_0)a_0+(n_0-n_1)a_1+\cdots
+(n_{l-2}-n_{l-1})a_{l-1}+n_{l-1}a_l-(k+1)v\leq 0,
$$
where $v=a_0+\frac{p}{q}(a_l-a_0)$. Since $\delta$ is arbitrary,
we deduce that $v\in P$ and the proof is
achieved by using Lemma \ref{POSS} {\em (i)}. \hfill $\Box$ 

\subsection{Characterization of the non-existence set $G$}\label{SECTG}
Phrased differently, Theorem \ref{MAINR} claims to each velocity
$v\in P$ corresponds a set of the parameter $c$ for which the only
existing fronts are those of velocity $\bar{v}(c)$ and their shape is
unique, the sets corresponding to different velocities being disjoint. We now
describe both these sets and the complement of their union, i.e.\ the
set $G$. 

Firstly if $\bar{v}(c)\in P\setminus D(\phi(0,.))$ then 
$$
\phi(x,\bar{v})<\phi(0,\bar{v})=c\quad \forall x<0,
$$
and the fronts exist. If $\bar{v}(c)\in D(\phi(0,.))$ since we have 
$$
\phi(0-0,\bar{v})=\phi(0,\bar{v}-0),
$$
then either $\phi(0-0,\bar{v})<c$ and the fronts exist, or 
$\phi(0-0,\bar{v})=c$. In the latter case, condition (\ref{EXIST}) is
satisfied or not according to $\bar{v}(c)\in P_1$ or $\bar{v}(c)\not\in P_1$
where 
$$
P_1=\left\{v\in \Bbb R\ : \ 0\in {\lclos{E(\phi(.,v))\setminus \{0\}}}\right\}.
$$
In this way, we conclude that for $h_1,h_2$ and $v\in P$ fixed, there
exists a set of the parameter $c$ given by
\begin{equation}\label{INQE}
\begin{array}{l}
c=\phi(0,v)\quad {\rm if}\quad v\in P\setminus D(\phi(0,.))\\
\phi(0,v-0)<c\leq \phi(0,v)\quad {\rm if}\quad v\in D(\phi(0,.))\\
c=\phi(0,v-0)\quad {\rm if}\quad v\in D(\phi(0,.))\cap P_1
\end{array}
\end{equation}
for which (only) the fronts of velocity $\bar{v}(c)=v$ exist.

In particular, if for a given velocity $v\in P$, it exists an interval
such that the selected velocity is $\bar{v}(c)=v$ for $c$ in this
interval, $v$ is said to correspond to a plateau. According to the
previous conclusion, a velocity $v$ corresponds to a plateau iff $v\in
D(\phi(0,.))$. Notice that Proposition \ref{CONV} applied to relation
(\ref{FRON}) leads to 
$$
D(\phi(0,.))=\bigcup_{k=0}^{\infty}\frac{D(h_2)++_kD(h_1)}{k+1}.
$$

Let us now give some general conditions for the couplings to satisfy
in order for $G$ (the set of $c$ values for which no front exist) 
to be countable and/or nowhere dense. According to the previous
reasoning, this set is given by $G=G_1\cup G_2$ where 
$$
G_1=\left\{\phi(0,v)\ :\ v\in E(\phi(0,.))\setminus P\right\}\quad
{\rm and}\quad G_2=\left\{\phi(0-0,v)\ :\ v\in D(\phi(0,.))\setminus P_1
\right\}.
$$
$G_2$ is countable since the discontinuity set of an increasing
function is \cite{Kolmo}. It is nowhere dense by the inequalities
(\ref{INQE}). Hence it remains to investigate these properties for
$G_1$. 
\begin{Pro}\label{GNUM}
$G$ is countable if one of the following conditions holds

\noindent
(i) $\#E(h_{1})\neq 1$

\noindent
(ii) $\#E(h_{1})=1$ and $E(h_{2})$ is bounded.
\end{Pro}
It is possible to find some examples in which $G$ is not countable. 

\noindent
{\sl Proof:} Firstly if $h_1=0$ then $\phi(x,v)=h_2(x+v)$. In this
case we obviously have $E(\phi(0,.))=P$ and thus
$G_1=\emptyset$. Assume now that $\#E(h_{1})=1$ and $E(h_{2})$ is
bounded. Then statement {\em (ii)} of Lemma \ref{POSS} implies
that $\# G_1\leq 1$. Finally if $\#E(h_{1})>1$ then
$(i_1,s_1)\setminus P$ is countable as follows from statement {\em
(i)} of Lemma \ref{POSS}. So is $\left(E(\phi(0,.))\setminus
P\right)\setminus [i_1,s_1]$ as it can be deduced from the assertions
{\em (ii)} and {\em (iii)} of the same lemma. \hfill $\Box$

To state conditions under which $G$ is nowhere dense in $(0,1)$, 
it is useful to introduce a notation and some
definitions. Given two real subsets $A$ and $B$, the notation $A\incinf
B$ has the following meaning
$$
A\incinf B\quad {\rm iff}\quad \left\{\begin{array}{l}
\sup A=+\infty \Rightarrow \sup B=+\infty\\
\inf A=-\infty \Rightarrow \inf B=-\infty
\end{array}\right.
$$
In particular, if $A$ is bounded no restriction is imposed on $B$ and
thus $A\incinf B$ for any $B\subset \Bbb R$. The second concept that
will be of use is the following.
\begin{Def}
A subset $A$ of $\Bbb R$ is commensurable if 
$$
\forall x_0,x_1,x_2\in A\quad x_2\neq x_0\Rightarrow
\frac{x_1-x_0}{x_2-x_0}\in \Bbb Q.
$$
$A$ is said to be incommensurable otherwise.
\end{Def}
\begin{Pro}\label{GDEN}
$G$ is nowhere dense if one of the following conditions is satisfied

\noindent
(i) $E(h_1)$ is incommensurable

\noindent
(ii) $h_1=0$

\noindent
(iii) $\#E(h_1)=1$ and $E(h_2)\incinf D(h_2)$

\noindent
(iv) $E(h_2)\cap R(E(h_1))\neq\emptyset$ and $E(h_2)\incinf D(h_2)\cup E(h_1)$

\noindent
(v) $D(h_2)\neq\emptyset$ and $E(h_2)\incinf D(h_2)\cup E(h_1)$.
\end{Pro}
Although many cases do not fit in these propositions, we can conclude
that $G$ is countable and nowhere dense in most of the physical
examples of piece-wise affine bistable extended mappings. Actually,
Propositions \ref{GNUM} and \ref{GDEN} cover the case of cellular
automata (i.e.\ $h_1=0$), the case of lattice dynamics (i.e.\ $D(h_2)=E(h_2)$)
(with finite range for $G$ to be countable) from which belong LDS and
CML, and more generally the
case $E(h_2)\subset E(h_1)$ that is expected to occur in the model
(\ref{GEND}) since we have 
$$
{\cal L}u+{\cal L}'F\circ u=\left({\cal L}+a{\cal
L}'\right)u+(1-a){\cal L}'H_c\circ u,
$$
and thus ${\cal L}+a{\cal L}'$ is expected to act at all the points
where ${\cal L}'$ acts. 

\noindent
{\sl Proof of Proposition \ref{GDEN}:} 
{\em (i)} If $E(h_1)$ is incommensurable then Proposition
\ref{ALLV} implies $(i_1,s_1)\subset P$. Now if $s_2=+\infty$ let 
$x_0,x_1,x_2\in E(h_1)$ be such that
$\frac{x_1-x_0}{x_2-x_0}\in\Bbb R\setminus\Bbb Q$. 
By using statement {\em (iii)}
of Lemma \ref{POSS} with $a=x_0$, $b=x_1$ and with $a=x_0$,
$b=x_2$ respectively, we conclude that $(s_1,+\infty)\subset P$. If
$s_2<+\infty$ then Lemma \ref{POSS} {\em (ii)} results in
$(s_1,s_2)\cap E(\phi(0,.))\setminus P=\emptyset$. One can proceed
similarly for the cases $i_2=-\infty$ and $i_2>-\infty$ to conclude
that $E(\phi (0,.))\setminus P$ has at most two points ($i_1$ and
$s_1$) and consequently $\#G_{1}\leq 2$. 

\noindent
If $h_1=0$ then $G_1=\emptyset$ (see the proof of Proposition
\ref{GNUM}) and {\em (ii)} follows. 

\noindent
Since we have already proved {\em (i)} and {\em (ii)}, 
we can now assume that $E(h_1)$ is
commensurable and non-empty. 
Hence it is countable and using Lemma \ref{LEDE}, we have $E(h_1)=D(h_1)$.
The proof of the remaining assertions is based on the following lemma proved in
Appendix \ref{TPROO}. 

\begin{Lem}\label{LDEN}
(i) If $D(h_2)\neq\emptyset$ and $E(h_1)=D(h_1)$,
then $G$ is nowhere dense in $\bar{v}^{-1}((i_1,s_1))$.

\noindent
(ii) If $\sup D(h_2)=+\infty$ and $a\in D(h_1)$, then $G$ is nowhere
dense in $\bar{v}^{-1}((a,+\infty ))$.

\noindent
(iii) If $\inf D(h_2)=-\infty$ and $a\in D(h_1)$, then $G$ is nowhere
dense in $\bar{v}^{-1}((-\infty ,a))$.
\end{Lem}
To achieve the proof of Proposition \ref{GDEN}, we are going to show
that under the conditions in {\em (iii)}, {\em (iv)} and {\em (v)}, 
$G_1$ is nowhere dense in each of the following intervals 
$\bar{v}^{-1}((i_2,i_1))$, $\bar{v}^{-1}((i_1,s_1))$ and 
$\bar{v}^{-1}((s_1,s_2))$, concluding that $G_1$ and consequently 
$G$ is nowhere dense in $(0,1)$.

\noindent
Firstly we prove that the condition $E(h_2)\incinf D(h_2)\cup E(h_1)$ 
implies that $G$ is nowhere dense in $\bar{v}^{-1}((s_1,s_2))$ by 
considering the following cases. If $\sup D(h_2)=+\infty $, then we use
Lemma \ref{LDEN} {\em (ii)}. If 
$\sup E(h_1)=+\infty $, then there is nothing to prove because 
$\bar{v}^{-1}((s_1,s_2))=\bar{v}^{-1}(\emptyset)=\emptyset$. If 
$\sup E(h_2)<+\infty $ Lemma \ref{POSS} statement {\em (ii)} results
in $\bar{v}^{-1}((s_1,s_2))\cap G_1=\emptyset $.

\noindent
Similarly $G$ is nowhere dense in $\bar{v}^{-1}((i_2,i_1))$.

\noindent
Now if $\#E(h_1)=1$ then $\bar{v}^{-1}((i_1,s_1))$ is empty, 
proving {\em (iii)}. If $E(h_2)\cap R(E(h_1))\neq \emptyset $ then by
Proposition \ref{ALLV}, $\bar{v}^{-1}((i_1,s_1))\cap G_1=\emptyset$
and {\em (iv)} is proved. Finally if $D(h_2)\neq\emptyset$ then by 
Lemma \ref{LDEN} {\em (i)}, $G$ is nowhere dense in 
$\bar{v}^{-1}((i_1,s_1))$ and {\em (v)} holds. \hfill $\Box $

\subsection{Continuity of the fronts velocity}
We now consider the dependence of the fronts velocity on the
couplings. In particular, we are going to show that
small variations of the couplings induce small variations of the
selected velocity. As couplings variations, we also would like to allow
for changes in the sets of increase points of the couplings 
$h_1$ and $h_2$. For instance,
this is the case when the planar fronts' direction varies 
in multi-dimensional CML (see Section \ref{CMLE}). We start
with a definition of a distance in ${\cal I}$. For $h,h'\in {\cal I}$ let 
$$
d(h,h')=\inf\left\{\epsilon>0\ :\ h(x-\epsilon)-\epsilon\leq h'(x)\leq
h(x+\epsilon)+ \epsilon\quad \forall x\in \Bbb R\right\}.
$$
One easily checks that $d(.,.)$ is a distance. For this
distance, the ball of radius $\epsilon$ centered at $h$ is the set of
functions for which the graphic lies in the band of
width $\epsilon$ in the direction of the line $y=-x$ around the graphic
of $h$. Similarly in the set of left continuous increasing
functions defined on $(0,1)$, we define $\tilde
d(v,v')$ as the infimum of the positive numbers $\epsilon$ such that 
$$
\left\{\begin{array}{ccl}
v(x-\epsilon)-\epsilon\leq v'(x)&{\rm if}& x\in (\epsilon,1)\\
v'(x)\leq v(x+\epsilon)+\epsilon&{\rm if}& x\in (0,1-\epsilon)
\end{array}\right. .
$$
From this definition one concludes that 
$\tilde d(.,.)$ is a distance and $\tilde d(.,.)\leq 1$. 
The continuity of the selected velocity in the subsequent
topology is an immediate consequence of the following statement. 
Let $\bar{v}$ (resp.\ $\bar{v}'$) be given by (\ref{BARV}) for the 
couplings $h_1$ and $h_2$ (resp.\ $h_1'$ and $h_2'$).
\begin{Pro}\label{CONTI}
For the couplings $h_1,h_1',h_2,h_2'$, let
$$m=max\{d(h_1,h_1'),d(h_2,h_2')\}.$$
Then 
$$
\tilde d(\bar{v},\bar{v}')\leq \frac{2m}{1-\|h_1'\|}.
$$
As a consequence, for any fixed $c\in\bar{v}^{-1}((i_1,s_1))$, 
$\bar{v}$ is a continuous function of the couplings in the Euclidean topology. 
\end{Pro}
{\sl Proof:} Let us firstly show the
continuous dependence of the shape in the topology induced by $d(.,.)$.
We notice that for any $\epsilon>d(h,h')$, the following holds
$$
H_{\epsilon}\ast h'-\epsilon \leq h\leq H_{-\epsilon}\ast h'+\epsilon .
$$
Hence by induction and by using the convolution properties, 
we obtain for any $k\geq 1$
$$
H_{k\epsilon}\ast h'^{\ast
k}-\epsilon\sum_{i=0}^{k-1}\|h\|^i\|h'\|^{k-1-i}\leq
h^{\ast k}\leq H_{-k\epsilon}\ast h'^{\ast k}+
\epsilon\sum_{i=0}^{k-1}\|h\|^i\|h'\|^{k-1-i}.
$$
Replacing these inequalities in the expression of $\phi$ and using
again the convolution properties result in the following 
$$
\forall \epsilon>m\quad 
\phi'(x,v-\epsilon)-\frac{2\epsilon}{1-\|h_1'\|}\leq \phi(x,v)\leq 
\phi'(x,v+\epsilon)+\frac{2\epsilon}{1-\|h_1'\|} ,
$$
where $\phi'$ is computed with $h_1'$ and $h_2'$. From the right
inequality we obtain 
$$
c-\frac{2\epsilon}{1-\|h_1'\|}\leq\phi(0,\bar{v}(c))
-\frac{2\epsilon }{1-\|h_1'\|}\leq\phi'(0,\bar{v}(c)+\epsilon),
$$
and hence by the definition of $\bar{v}$, it follows 
$$
\bar{v}'\left(c-\frac{2\epsilon }{1-\|h_1'\|}\right)\leq\bar{v}(c)+\epsilon.
$$
Similarly we have
$$
c-\frac{2\epsilon}{1-\|h_1'\|}\leq\phi'(0,\bar{v}'(c))
-\frac{2\epsilon}{1-\|h_1'\|}\leq\phi(0,\bar{v}'(c)+\epsilon),
$$
and by the definition of $\bar{v}$ and by replacing $c$ by 
$c+\frac{2\epsilon}{1-\|h_1'\|}$, we obtain
$$
\bar{v}(c)\leq\bar{v}'\left(c+\frac{2\epsilon}{1-\|h_1'\|}\right)+\epsilon.
$$
Hence $\tilde d(\bar{v},\bar{v}')\leq\frac{2\epsilon}{1-\|h_1'\|}$ 
and the statement is obtained by choosing $\epsilon$ arbitrarily close
to $m$.

Finally, we assume that $\bar{v}(c)\in (i_1,s_1)$. We have shown that 
$$
\bar{v}\left(c-\frac{2\epsilon}{1-\|h_1'\|}\right)
-\frac{2\epsilon}{1-\|h_1'\|}\leq\bar{v}'(c)\leq
\bar{v}\left(c+\frac{2\epsilon}{1-\|h_1'\|}\right)
+\frac{2\epsilon}{1-\|h_1'\|},
$$
for $\epsilon>m$. Thus from the continuity of $\bar{v}(.)$ in 
$\bar{v}^{-1}((i_1,s_1))$, we obtain for $c$ fixed
$$
\lim_{m\rightarrow 0}\bar{v}'(c)=\bar{v}(c).
$$
\hfill $\Box $

\subsection{The anti-fronts}
To conclude the study of the existence of travelling wave
interfaces in our dynamical system, we mention that the results about 
the fronts can be employed to deduce the existence of anti-fronts and
the uniqueness of their shape. An anti-front is a travelling wave for
which the shape $\phi^{{\rm anti}}$ obeys the condition 
$$
H_c\circ \phi^{{\rm anti}}(x)=H_0(-x)\quad\forall x\in \Bbb R.
$$
By defining $\phi^{\rm a}(x)=\phi^{{\rm anti}}(-x)$ and by using the
shape equation, one can easily prove the following statement. 
\begin{Pro}
The dynamical system (\ref{DEFD2}) has anti-fronts of
velocity $v$ and shape $\phi^{{\rm anti}}$ iff there exist fronts of
velocity $-v$ and shape $\phi^{\rm a}$ for the dynamics given by
\[
u^{t+1}=h_1^{{\rm a}}*u^t+h_2^{{\rm a}}*H_c\circ u^t,
\]
where $h_i^{{\rm a}}(x)=\|h_i\|-h_i(-x-0)$ for $i=1,2$. 
\end{Pro}
Moreover it is possible to relate $\bar{v}^{{\rm anti}}(c)$ 
the velocity of anti-fronts to the fronts velocity for
the same couplings, i.e.\ those represented by $h_1$ and $h_2$, in the
following way
$$
\bar{v}^{{\rm anti}}(c)=\bar{v}(1-c+0).
$$
Hence in general the selected velocities of fronts and of
anti-fronts, as well as the corresponding 
non-existence sets $G$ and $G^{\rm anti}$,
are different.

\section{Velocity of interfaces}
In this section, an also important result, when one has in mind 
physical applications of bistable extended mappings, 
is presented. In fact we are going to prove
that the fronts velocity $\bar{v}$ is actually the propagation
velocity for any orbit of (\ref{DEFD2}) composed at each time of a
configuration linking two different phases, namely a $c$-interface.

\begin{Def}
A $c$-interface is a function $u\in {\cal M}$ such that 
$$
\exists J_{\rm i},J_{\rm s}\in \Bbb R\ :
\left\{\begin{array}{ccl}
u(x)<c&{\rm if}&x<J_{\rm i}\\
u(x)\geq c&{\rm if}&x>  J_{\rm s}
\end{array}\right. .
$$
The set of $c$-interfaces is denoted by ${\cal J}^c$ and $J_{\rm i}(u)$ and
$J_{\rm s}(u)$ are the functions from ${\cal J}^c$ to $\Bbb R$ defined by
$$
J_{\rm i}(u)=\inf\{x\ :\ u(x)\geq c\} \quad {\rm and}\quad 
J_{\rm s}(u)=\sup\{x\ :\ u(x)<c\}.
$$
\end{Def}
The following proposition guarantees that the image under the dynamics
(\ref{DEFD2}) of a $c$-interface is a $c$-interface. 
\begin{Pro}\label{INTFOREV}
If $c\in (0,1)$ and $u\in {\cal J}^c$ then 
$h_1*u+h_2*H_c\circ u\in {\cal J}^c$.
\end{Pro}
This statement is still true for $c=1$ if $v_{{\rm max}}<+\infty$. 

\noindent
{\sl Proof:} Let $u^1=h_1*u+h_2*H_c\circ u$ and $\tilde{u}$ be defined by
$$
\tilde{u}(x)=\left\{\begin{array}{ccl}
u(x)&{\rm if}&x\leq  J_{\rm s}(u)\\
c&{\rm if}&x>  J_{\rm s}(u)
\end{array}\right.
$$
We have $\tilde{u}\leq u$. By using the convolution
properties, we then obtain
$$
\liminf_{x\rightarrow +\infty}u^1(x)\geq c\|h_1\|+\|h_2\|>c,
$$
since $\|h_1\|=1-\|h_2\|$ and $c<1$. Similarly, one proves 
$$
\limsup_{x\rightarrow -\infty}u^1(x)\leq c\|h_1\|<c.
$$
\hfill $\Box $

\noindent
Hence the terminology $c$-interface orbit $\{u^t\}_{t\in \Bbb N}$ is 
justified. As it has been pointed out in the case of CML \cite{Coutinho2}, 
these orbits can evolve towards different attractors. So a general
common (asymptotic) stability result (in the sense of Lyapunov) in ${\cal J}^c$
cannot hold. However the $c$-interface orbits 
all have a common dynamical characteristic which is the same propagation
velocity.  
\begin{Thm}\label{VINTE1} 
If $c$ is a point of continuity of the function $\bar{v}$ and 
$\{u^t\}_{t\in \Bbb N}$ is a $c$-interface orbit, then its velocity
exists and is given by $\bar{v}(c)$, that is to say 
$$
\lim_{t\rightarrow +\infty}\frac{J_{\rm i}(u^t)}{t}=
\lim_{t\rightarrow +\infty}\frac{J_{\rm s}(u^t)}{t}=\bar{v}(c).
$$
\end{Thm}
The result cannot hold if $c$ is a discontinuity point of
$\bar{v}$ because in this situation, the dynamical system has (also) 
ghost fronts of velocity $\bar{v}(c+0)$ which attract some
initial conditions. The proof of Theorem \ref{VINTE1} 
is based on the following proposition.
\begin{Pro}
For any fixed $c\in (0,1)$ we have $\forall u^0\in {\cal J}^c$ 
$\forall \epsilon >0$ $\exists \sigma_0,\sigma_1\in \Bbb R$ such that
$$
\sigma_0+t(\bar{v}(c)-\epsilon)\leq J_{\rm i}(u^t)\leq 
J_{\rm s}(u^t)\leq \sigma_1+t\bar{v}(c+\epsilon).
$$
\end{Pro}
Clearly if $v_{\rm max}<+\infty$, we have a similar claim for $c=1$.

\noindent
{\sl Proof:} Given $u^0\in {\cal J}^c$, by iterating the dynamics
(\ref{DEFD}) we obtain for $t\geq 1$
$$
u^t=h_1^{*t}*u^0+\sum_{k=0}^{t-1}h_2*h_1^{*k}*H_c\circ u^{t-1-k}.
$$
Since $\{u^t\}_{t\in\Bbb N}$ is composed of $c$-interfaces, the
convolution properties and the definitions of $J_{\rm i}$ and of
$J_{\rm s}$ lead to the following inequalities 
$$
h_1^{*t}*u^0+\sum_{k=0}^{t-1}h_2*h_1^{*k}*H^0_{J_{\rm s}(u^{t-1-k})}
\leq u^t\leq 
h_1^{*t}*u^0+\sum_{k=0}^{t-1}h_2*h_1^{*k}*H_{J_{\rm i}(u^{t-1-k})},
$$
where $H^0_{\omega}(x)=H_{\omega}(x-0)$ $\forall \omega,x\in\Bbb R$. 
Now given $\epsilon >0$ let
$$
\Delta_0=c-\phi(0,\bar{v}(c)-\epsilon)\quad {\rm and}\quad 
\Delta_1=\phi(0,\bar{v}(c+\epsilon))-c
$$
By the definition of $\bar{v}(c)$, we have $\Delta_0>0$ and $\Delta_1\geq
\epsilon>0$. Let then $t_0,t_1\in \Bbb N$ be such that
$$
\|h_1\|^{t_0}\|u^0\|<\Delta_0\quad {\rm and}\quad 
\|h_1\|^{t_1}(\|u^0\|+1)<\Delta_1,
$$
and $\sigma_0,\sigma_1\in \Bbb R$ be defined by
$$
\sigma_0=\min_{0\leq k\leq t_0-1}\left(J_{\rm
i}(u^k)-k(\bar{v}(c)-\epsilon)\right) \quad {\rm and}\quad 
\sigma_1=\max_{0\leq k\leq t_1-1}\left(J_{\rm
s}(u^k)-k\bar{v}(c+\epsilon)\right).
$$
We are going to prove by induction that
\begin{equation}\label{IND}
\sigma_0+k(\bar{v}(c)-\epsilon)\leq J_{\rm i}(u^k)\leq J_{\rm s}(u^k)\leq 
\sigma_1+k\bar{v}(c+\epsilon)\quad \forall k\in \Bbb N.
\end{equation}
By construction the left (resp.\ right) inequality is true for 
$0\leq k\leq t_0-1$ (resp.\ $0\leq k\leq t_1-1$). Assume
the left (resp.\ right) inequality in (\ref{IND}) holds for 
$0\leq k\leq t-1$ with $t\geq t_0$ (resp.\ $t\geq t_1$). It comes out
that 
$$
u^t(x)\leq \|h_1\|^{t_0}\|u^0\|+
\sum_{k=0}^{t-1}h_2*h_1^{*k}*H_0(x-\sigma_0-(t-1-k)(\bar{v}(c)-\epsilon))\quad
\forall x\in\Bbb R,
$$
and 
$$
u^t(x)\geq -\|h_1\|^{t_1}\|u^0\|
+\sum_{k=0}^{t-1}h_2*h_1^{*k}*H^0_0(x-\sigma_1-(t-1-k)\bar{v}(c+\epsilon))\quad
\forall x\in\Bbb R.
$$
Hence using the definition of the function $\phi(z,v)$ we have
$$
u^t(x)\leq \|h_1\|^{t_0}\|u^0\|+
\phi(x-\sigma_0-t(\bar{v}(c)-\epsilon),\bar{v}(c)-\epsilon)\quad
\forall x\in\Bbb R,
$$
and for any $\delta>0$
$$
u^t(x)\geq -\|h_1\|^{t_1}(\|u^0\|+1)+
\phi(x-\sigma_1-\delta-t\bar{v}(c+\epsilon),\bar{v}(c+\epsilon ))\quad
\forall x\in\Bbb R,
$$
the latter being obtained by using the relation
$\left\|\sum\limits_{k=t_1}^\infty
h_2*h_1^{*k}\right\|=\|h_1\|^{t_1}$. Consequently if 
$x\leq\sigma_0+t(\bar{v}(c)-\epsilon)$ we have
$$
u^t(x)\leq \|h_1\|^{t_0}\|u^0\|+\phi(0,\bar{v}(c)-\epsilon)<c,
$$
in other terms $\sigma_0+t(\bar{v}(c)-\epsilon) \leq J_{\rm
i}(u^t)$. If $x\geq \sigma_1+\delta+t\bar{v}(c+\epsilon )$ then
$$
u^t(x)\geq -\|h_1\|^{t_1}\left(\|u^0\|+1\right)+\phi(0,\bar{v}(c+\epsilon))>c,
$$
proving the right inequality 
$J_{\rm s}(u^t)\leq\sigma_1+t\bar{v}(c+\epsilon )$.\hfill $\Box $

\noindent 
{\sl Proof of theorem \ref{VINTE1}:} 
From the last proposition, it is deduced the following inequalities
$$
\forall\epsilon>0\quad\bar{v}(c)-\epsilon\leq\liminf_{t\rightarrow +\infty}
\frac{J_{\rm i}(u^t)}t\leq\limsup_{t\rightarrow +\infty}
\frac{J_{\rm s}(u^t)}t\leq\bar{v}(c+\epsilon),
$$
and thus 
$$
\bar{v}(c)\leq\liminf_{t\rightarrow +\infty}\frac{J_{\rm i}(u^t)}t\leq
\limsup_{t\rightarrow +\infty}\frac{J_{\rm s}(u^t)}t\leq \bar{v}(c+0).
$$
Since $J_{{\rm i}}(u^t)\leq J_{{\rm s}}(u^t)$ we obtain the existence of the
limits if $\bar{v}(c)=\bar{v}(c+0)$.\hfill $\Box $

\section{Examples}
In order to illustrate the results obtained in the general framework,
we now present three examples. One focuses on discrete couplings, the
second deals with continuous couplings, and we thirdly
show how planar fronts in
multi-dimensional CML enter in the general framework.\footnote{In this
section, we will refer both to the set $P$ of possible velocities and
to $G$ the set of values of $c$ for which no front exists. (See proof
of Theorem \ref{MAINR})} 

\subsection{One-dimensional Lattice Dynamical System}\label{LDSE}
In this first example, we choose the couplings as
acting on a one-dimensional lattice and as being of finite range.
The interest here resides in emphazing on two characteristics. One is
the absence of some rational velocities
in $(i_1,s_1)$ and the other is the structure of $P$ in
$(i_2,s_2)\setminus (i_1,s_1)$ when the set of increase points of
$h_2$ is bounded. 

In order to have these both characteristics, we set for $a\in (0,1)$
and $\epsilon \in (0,1)$
\begin{equation}\label{DISEX}
L_1=\frac{a}{2}\left(\sigma^1+\sigma^{-1}\right)\quad 
{\rm and}\quad 
L_2=(1-a)\left((1-\epsilon){\rm Id} +\frac{\epsilon}{2}
\left(\sigma^2+\sigma^{-2}\right)\right),
\end{equation}
where ${\rm Id}$ is the identity on ${\cal M}$. It is immediate to
verify that these obey the linearity, continuity, positivity,
s-homogeneity and normalization conditions. It is also easy to see
that $D(h_1)=E(h_1)=\{-1,1\}$ and $D(h_2)=E(h_2)=\{-2,0,2\}$ and
consequently $E(h_2)\cap R(E(h_1))=\emptyset$. Therefore,
Proposition \ref{ALLV} does not apply and
some rational velocities are expected to be absent from $P$. The results are
contained in the following statement. In the remaining of this
section, $p$ and $q$ denote two co-prime integers. 
\begin{Pro}\label{EXA1}
For the couplings given by (21) we have $P\subset [-2,2]$ and

\noindent
(i) the velocity $v\in [-1,1]$ belongs to $P$ iff either 
$v$ is irrational or $v=\frac{p}{q}$ with $p+q$ odd,

\noindent
(ii) the velocity $v\in [1,2]$ (resp.\ $[-2,-1]$) belongs to $P$ iff it
exists $k\in \Bbb N$ such that $v=1+\frac{1}{k+1}$ (resp.\
$v=-1-\frac{1}{k+1}$). 
\end{Pro}
Notice that the set 
$$[-1,1]\setminus P=\left\{v\in[-1,1]\ :\ v=\frac{p}{q}, \ p+q \ {\rm even}
\right\}$$ 
is dense in $[-1,1]$ but $G$ is nowhere dense.

We have plotted on Figure 2 the selected velocity 
$\bar{v}(c)$ versus $c\in[\frac{1}{2},1]$, the function
being symmetric, i.e.\ $\bar{v}(c)=-\bar{v}(1-c+0)$. This function is
a Devil's staircase in $[\bar{v}^{-1}(-1),\bar{v}^{-1}(1)]$ and a
classical staircase in the complementary intervals.

\begin{figure}
\epsfxsize=12truecm 
\centerline{\epsfbox{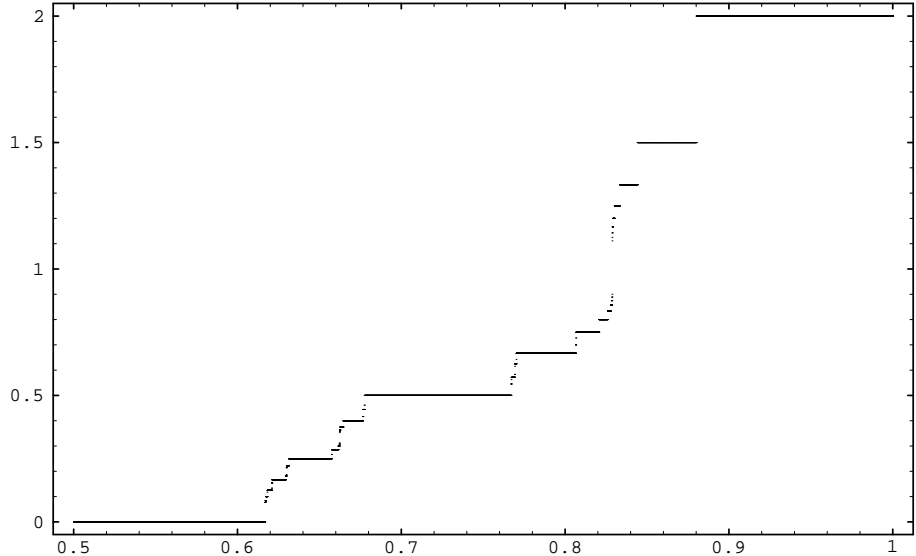}}
\caption{The selected velocity $\bar{v}(c)$ versus $c$ for the
one-dimensional lattice dynamical system example ($a=\epsilon=0.6$).}
\end{figure}

\noindent
{\sl Proof of Proposition \ref{EXA1}:} {\em (i)} By a direct
calculation, we obtain
$$
\phi(x,v)=(1-a)\sum_{k=0}^{\infty}\sum_{n\in\Bbb Z}\left((1-\epsilon)
l_n^k+\frac{\epsilon}{2}(l_{n-2}^k +l_{n+2}^k)\right)H_{n-v(k+1)}(x)
\quad \forall x\in \Bbb R,
$$
where the coefficients $l_n^k$ are the entries of the powers $L_1^k$. These
coefficients are known explicitly \cite{Coutinho96}. In particular
they obey the following property
\begin{equation}\label{COEFF}
l_n^k\neq 0\quad {\rm iff}\quad |n|\leq k \ {\rm and}\
\frac{n+k}{2}\in \Bbb Z.
\end{equation}
On one hand all the irrational velocities $[-1,1]$ are
possible as given by Lemma \ref{POSS} {\em (i)}. 
On the other hand we have 
$$
\phi(x,\frac{p}{q})=\phi(0-0,\frac{p}{q})=\phi(0,\frac{p}{q}-0)\quad
{\rm for}\quad x\in \left[-\frac{1}{q},0\right),
$$ and thus the proof of Theorem \ref{MAINR} tells that  
$\frac{p}{q}\in [-1,1]$ belongs to $P$ iff
$\phi(0,\frac{p}{q}-0)<\phi(0,\frac{p}{q})$. Using the definition of
the Heaviside function, we have
$$
\phi(0,\frac{p}{q})-\phi(0,\frac{p}{q}-0)=(1-a)\sum_{k=1}^{\infty}
\left((1-\epsilon)l_{kp}^{kq-1}+\frac{\epsilon}{2}(l_{kp-2}^{kq-1} +
l_{kp+2}^{kq-1})\right)
$$
Taking into account relation (\ref{COEFF}), it turns out that 
$$
\phi(0,\frac{p}{q})=\phi(0,\frac{p}{q}-0)\quad {\rm iff}\quad \forall
k\geq 1, \ \exists j\in \Bbb N\ : \ k(p+q)=2j,
$$
from which we conclude that $\frac{p}{q}\in P$ iff $p+q$ is odd.

\noindent
{\em (ii)} is easily deduced from the proof of statement 
{\em (ii)} Lemma \ref{POSS}. \hfill $\Box $

\subsection{Continuous example}
As a continuous couplings example, we consider for $a\in (0,1)$, 
$L_1=aL$ and $L_2=(1-a)L$ where $L$ is now the continuous diffusion 
operator expressed in the integral formulation, i.e.\ the corresponding
function $h$ is defined by
$$
h(x)=\int_{-\infty}^x e^{-\pi y^2}dy,
$$
so that the couplings obey the linearity, continuity, positivity, 
s-homogeneity and normalization conditions. In this case we have
$E(h_1)=E(h_2)=\Bbb R$ and $D(h_1)=D(h_2)=\emptyset$ thus $P=\Bbb R$
and $G$ is empty. We conclude that for any $a,c\in (0,1)$ it 
exists a unique front shape. The corresponding velocity is a
continuous function of (a,c) with range $\Bbb R$. 

The front velocity $\bar{v}(c)$ in this continuous model, restricted
to the values in $[0,1]$, is shown on Figure 3. Notice that $\bar{v}(.)$ is an
odd function.
\begin{figure}
\epsfxsize=12truecm 
\centerline{\epsfbox{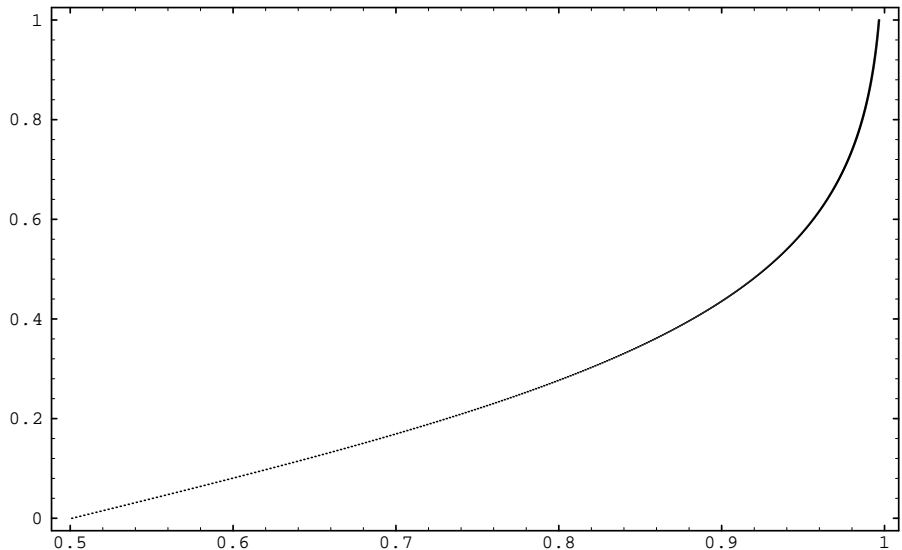}}
\caption{The selected velocity $\bar{v}(c)$ versus $c$ for the
continuous coupling ($a=0.4$).}
\end{figure}

\subsection{Multi-dimensional Coupled Map Lattices}\label{CMLE}
We now turn to the application of the previous results to the dynamics
of planar fronts in discrete
bistable extended systems defined on a $d$-dimensional lattice, by
considering a simple situation, that of homogeneous CML. However a
similar analysis can be applied, without supplementary difficulties,
 to more general multi-dimensional extended mappings. After
defining these systems, we shall see how the homogeneity
implies the reduction of the dynamics, for the orbits
having equal components on $(d-1)$-dimensional planes, to a
one-dimensional extended mapping. The
existence of planar fronts in these multi-dimensional models then directly
follows from this reduction and from Theorem \ref{MAINR}. Finally, the
monotonicity of the dynamics allows to conclude about the velocity of
interfaces, even when their dynamics does not reduce to a
one-dimensional extended mapping. 

\subsubsection{Definitions}
Multi-dimensional CML are introduced following the ideas of
\cite{Kaneko}. For $d\in \Bbb N$, $d >0$, let 
$$
{\cal M}'=\left\{ u=\{u_s\}_{s\in \Bbb Z^d},\ u_s\in \Bbb R\ :\ 
\|u\|_{\infty}<+\infty \right\}, 
$$
be the phase space under consideration. A homogeneous $d$-dimensional
CML is a discrete time dynamical system for which each time iteration of
an orbit $\{u^t\}$ of elements of ${\cal M}'$ is
given by the following relation 
\begin{equation}\label{MULTI}
u^{t+1}=LF\circ u^t.
\end{equation}
Here the mapping $F$ is a direct product of one-dimensional
identical mappings, i.e.
$$
\left(F\circ u\right)_s=f(u_s)\quad \forall s\in {\Bbb Z}^d,
$$
where $f$ maps $\Bbb R$ into itself. The coupling $L$ 
is required to be a linear, continuous, positive and
s-homogeneous operator of ${\cal M}'$ into itself. Its explicit
expression is then
$$
\left(Lu\right)_s=\sum_{n\in {\Bbb Z}^d}l_nu_{s-n}\quad \forall
s\in {\Bbb Z}^d,
$$
where the real coefficients $l_n\geq 0$ obey the normalization condition
$$
\sum_{n\in {\Bbb Z}^d}l_n=1.
$$

A point $s$ in  the multi-dimensional lattice ${\Bbb Z}^d$ is denoted by
$\{s_i\}_{1\leq i\leq d}$. The difference of two elements in ${\Bbb
Z}^d$ is to be understood as the component-wise difference. Moreover we are
going to use the Euclidean inner product in ${\Bbb R}^d$, i.e.
$$
r.s=\sum_{i=1}^dr_is_i\quad \forall r,s\in {\Bbb R}^d.
$$

\subsubsection{One-dimensional orbits}
Multi-dimensional homogeneous CML satisfy any
translational invariance symmetry along each lattice axis. To each
of these symmetries corresponds a reduction of the dimension of the
lattice supporting the dynamics. Hence by combining $d-1$ linearly independent
translations, the resulting orbit's dynamics is governed by a 
one-dimensional extended mapping, as we describe now. 

Let $k\in {\Bbb R}^d$ be such that $k.k=1$, a fixed direction. By
induction and by using the coupling homogeneity, one shows 
that for any orbit $\{u^t\}_{t\in \Bbb N}$ generated by the dynamics
(\ref{MULTI}), if the property
\begin{equation}\label{PROT}
\forall r,s\in {\Bbb Z}^d \ :\ k.r=k.s\Rightarrow u^t_r=u^t_s,
\end{equation}
holds for $t=0$, then it is satisfied for any $t\geq 0$. In this
case\footnote{This restriction is not necessary if $k$ is an
irrational direction, i.e.\ if $k.s=0$ implies $s=0$.}
the time evolution reduces to that of a one-dimensional discrete extended
mapping. Actually
by introducing the reduced coordinate $\omega=k.s$ and the
corresponding component $z_{\omega}^t=u_s^t$, we obtain the
component-wise expression for the reduced dynamics 
\begin{equation}\label{REDU}
z^{t+1}_{\omega }=\sum_{\nu \in Z(k)}\lambda_{\nu}
f\left(z^t_{\omega-\nu}\right)\quad \forall \omega\in Z(k),
\end{equation}
where 
$$
Z(k)=\left\{\omega\in \Bbb R\ :\ \omega =k.s,\ s\in {\Bbb
Z}^d\right\}\quad {\rm and}\quad \lambda_{\omega }=\sum_{n\ :\
k.n=\omega}l_n \quad \forall \omega\in Z(k).
$$
Similarly as in Remark \ref{RESR}, if $\forall i,j$
$\frac{k_i}{k_j}\in \Bbb Q$ then it exists $m\in \Bbb R^+$ such that
$Z(k)=m\Bbb Z$ and if $\exists i,j$ such that $\frac{k_i}{k_j}\in\Bbb
R\setminus \Bbb Q$ then $Z(k)$ is dense in $\Bbb R$. Furthermore 
for a given multi-dimensional coupling $L$, the 
one-dimensional (discrete) reduced coupling $\Lambda$ defined by
$$
\left(\Lambda z\right)_{\omega }=\sum_{\nu \in Z(k)}\lambda_{\nu }
z_{\omega-\nu},
$$
and thus the dynamical properties of the corresponding one-dimensional
orbits of (\ref{MULTI}), depend on the direction $k$.
From now on we choose as a local map, the piece-wise affine model
(\ref{LOC}). The reduced system (\ref{REDU}) then turns out to be a
particular case of the system (\ref{DEFD2}) with the coupling
functions given by $h_1=ah_k$ and $h_2=(1-a)h_k$
where
\begin{equation}\label{HMULT}
h_k(\omega)=\sum_{\nu \in Z(k)}\lambda_{\nu }H_0(\omega-\nu)=
\sum_{n\in \Bbb Z^d}l_nH_0(\omega-k.n).
\end{equation}

\subsubsection{The planar fronts}
\begin{Def}
A planar front of velocity $v$
in the direction $k$ is an orbit of (\ref{MULTI}) defined by 
$$
u^t_s=\phi(k.s-vt-x_0)\quad \forall t\in {\Bbb Z}, s\in {\Bbb Z}^d,
$$
where $x_0\in\Bbb R$ and the shape $\phi\in {\cal M}$ and obeys the
condition (\ref{EXIST}).
\end{Def}
As for fronts, the existence of planar
fronts only depend on the existence of their shape, and is therefore
independent of the translation parameter $x_0$.

Since the planar fronts are
invariant under translations orthogonal to their direction, we can
apply the previous reduction to obtain the equivalence between the
planar fronts of (\ref{MULTI}) and the fronts of the corresponding reduced
system (\ref{REDU}). The planar front existence in piece-wise affine bistable
$d$-dimensional CML then follows from Theorem \ref{MAINR} and the 
propositions of Section \ref{MAINSE}. Let us
denote by $i_k=\inf\left\{\omega\in Z(k)\ :\ \lambda_{\omega
}>0\right\}$ and by $s_k=\sup \left\{\omega\in Z(k)\ :\ \lambda_{\omega
}>0\right\}$. 
\begin{Cor}
Given $a,L$ and $k$, there is a countable nowhere dense set $G$ such
that, for any $c\in (0,1]\setminus G$, there exist planar
fronts in the system (21) with a unique shape. The corresponding
velocity is a continuous function of $(a,c,k,L)$ with range $[i_k,s_k]$.
\end{Cor}
In particular for a given coupling, the selected velocity as well
as the extreme velocities and the plateaus, depend continuously on
the planar fronts' orientation.

As a sketch of proof, one firstly notices that from (\ref{REDU}) it
follows the relations 
$E(h_1)=D(h_1)=E(h_2)=D(h_2)\subset Z(k)\cap [i_k,s_k]$ and hence
$P=E(\phi(0,.))=[i_k,s_k]$ and $G=G_2$. This allows to state the
existence following the existence's proof in the general case. The 
continuity versus $k$ and $L$ is a consequence of Proposition
\ref{CONTI} and of the fact that
small variations of $k$ and of $L$ induce small variations of the
reduced coupling $\Lambda$ for the distance $d(.,.)$ as can be easily
seen from (\ref{HMULT}). 

\subsubsection{Velocity of interfaces}
Similarly as in the one-dimensional case, the subsequent question to
the existence of fronts is the velocity of interfaces in
multi-dimensional CML. Naturally their definition must take into
account the orientation.
\begin{Def}
A $(k,c)$-interface is a configuration $u\in {\cal M}^{\prime }$ such that
\[
\exists J_{k{\rm i}},J_{k{\rm s}}\in {\fam\msbfam\relax R}\ :\left\{ 
\begin{array}{ccl}
u_s<c & {\rm if} & s.k<J_{k{\rm i}} \\ 
u_s\geq c & {\rm if} & s.k>J_{k{\rm s}}
\end{array}
\right. .
\]
The set of $(k,c)$-interfaces is denoted by ${\cal J}_k^c$ and 
$J_{k{\rm i}}(u)$ and $J_{k{\rm s}}(u)$ are the functions from 
${\cal J}_k^c$ to $\Bbb R$ defined by 
\[
J_{k{\rm i}}(u)=\inf \{s.k\ :\ u_s\geq c\}\quad {\rm and}\quad J_{k{\rm s}
}(u)=\sup \{s.k\ :\ u_s<c\}.
\]
\end{Def}

To each $(k,c)$-interface $u$ we can associate two $(k,c)$-interfaces
$\underline{u}$ and $\overline{u}$ satisfying relation (\ref{PROT})
and defined for any $s\in\Bbb Z^d$ by
\[
\underline{u}_s=\inf \{u_n\ :\ n.k=s.k\}\quad {\rm and}\quad
\overline{u}_s=\sup \{u_n\ :\ n.k=s.k\}. 
\]
We have $\underline{u}_s\leq u_s\leq \overline{u}_s$ $\forall
s\in\Bbb Z^d$ and then $J_{k{\rm i}}(\overline{u})\leq J_{k{\rm
i}}(u)\leq J_{k{\rm i}}(\underline{u})$ and $J_{k{\rm
s}}(\overline{u})\leq J_{k{\rm s}}(u)\leq J_{k{\rm
s}}(\underline{u})$. For the configurations $\underline{u}$ and
$\overline{u}$ the dynamics can be reduced to a one-dimensional 
extended mapping as was done for planar fronts. Consequently we can apply 
Proposition \ref{INTFOREV}, Theorem \ref{VINTE1} and the order
preservation of the dynamics to conclude that a $(k,c)$-interface 
initial condition $u^0$ generates a $(k,c)$-interface orbit 
$\{u^t\}_{t\in \Bbb N}$ and the following statement holds. 
\begin{Cor}
For any $u^0\in {\cal J}_k^c$ the following limits exist and do not depend
on the particular choice of $u^0$ 
$$
\lim_{t\rightarrow +\infty }\frac{J_{k{\rm i}}(u^t)}t=\lim_{t\rightarrow
+\infty }\frac{J_{k{\rm s}}(u^t)}t=\bar{v}_k,
$$
where $\bar{v}_k$ is the velocity of the planar front in the direction
$k$.
\end{Cor}

\section{Other fronts and general bistable extended mappings}\label{OMODEL}
Starting from the knowledge of the fronts dynamics in piece-wise
affine bistable extended mappings with homogeneous linear couplings,
one can extend the results to other kinds of fronts and to 
more general systems in which one or many
of the previous conditions are relaxed. Here we firstly consider
changes in the local map and then we allow for nonlinear couplings.

Let us recall that a fronts velocity $v$ corresponds to a plateau
iff it is a discontinuity point of the function $\phi(0,.)$. In this
case according to (\ref{INQE}), the interval corresponding to this
plateau is given by $I_v=(\phi(0-0,v),\phi(0,v)]$, the left bound being
included if $v\in P_1$. In the following $I_v^0$ denotes the interior
of this interval. 

In the interior of a plateau and without changing the couplings, 
the existence of fronts is claimed in 
systems defined by (\ref{GEND}) but with now
a $C^{\infty}$ bistable local map. Actually inside the plateaus, the 
distance from the front shape to the discontinuity $c$ is positive. 
In this case one can modify the local map in a
neighborhood $V$ of $c$ where the fronts never go, i.e.\ $V\subset
I_v^0$, without modifying these orbits. The existence of fronts
in some bistable extended mappings with a $C^{\infty}$ local map then
directly follows from this observation. Similarly it is possible to
extend the result on the velocity of interfaces to these mappings. 

Condition (\ref{EXIST}) in the front definition imposes the shape to
be a right continuous function as it is claimed in
relation (\ref{INCR}). However one can ask about the existence of
fronts of a different kind, namely fronts having a left continuous shape
satisfying instead of (\ref{EXIST}), the condition 
$$
\left\{\begin{array}{ccl}
\tilde\phi(x)<c&{\rm if}&x\leq 0\\
\tilde\phi(x)\geq c&{\rm if}&x>0
\end{array}\right.
$$
Obviously, if such fronts exist, we have 
$$
\tilde\phi(0)=\phi(x-0),
$$
where $\phi$ is given by (\ref{FRON}). Consequently according to
Theorem \ref{MAINR}, the dynamical 
system (\ref{DEFD2}) admits fronts with a left continuous shape
iff it has fronts with a right continuous shape $\phi$ which does not
touch $c$ (i.e.\ $\phi(0)\neq
c$). The latter occurs when the velocity belongs to the interior of a plateau. 
Moreover following the previous argument, fronts with a left
continuous shape also exist in a larger class of bistable extended
mappings for which smooth local maps are allowed. 

Still in $I_v^0$ for $v\in D(\phi(0,.))$, 
the local map can be modified (now only) at $c$ in order to obtain
a third kind of fronts still with the same velocity
$v$ but which will be unstable (in the sense of
Lyapunov).\footnote{Notice that Theorem \ref{VINTE1} imposes all the fronts
defined here to propagate with the same velocity.}
One has to consider the
following definition of the Heaviside function instead of the one used
up to now. 
$$
H^r_0(x)=\left\{\begin{array}{ccl}
0&if&x<0\\
r&if&x=0\\
1&if&x>0
\end{array}\right.
$$
for $0\leq r\leq 1$. The corresponding front shape follows from the
introduction of this function in condition (\ref{EXIST}) and from solving
the subsequent shape equation. The solution $\phi'$ now depends
continuously on $r$ and we have, writing explicitly the dependence on
$r$, 
$$
\phi'(0-0,v,1)=\phi'(0,v,0)<\phi'(0,v,r)<\phi'(0,v,1)=\phi(0,v).
$$
Hence given $c$ inside a plateau, 
the front shape $\phi'$ exists if $r$ is such that 
$$
\phi'(0,v,r)=c\quad {\rm and}\quad f(c)=ac+r(1-a).
$$
It is unstable since any small enough positive (resp.\ negative)
perturbation of $\phi'$ at 0 belongs 
to the basin of attraction of the
front with shape $\phi$ (resp.\ $\tilde\phi$). Actually one can show
the linear stability of the latter fronts inside the plateaus.
For $r=1$, $\phi'$ coincide with $\phi$ whereas for $r=0$ 
it is nothing but $\tilde\phi$. 
Moreover the extension from discontinuous to
$C^{\infty}$ local map can still be done inside a plateau by requiring
$f(c)=ac+r(1-a)$. The front shape built with the
present definition of the Heaviside function will be stable if 
$|f'(c)|<1$ and unstable if $|f'(c)|>1$.

According to these results, we conjecture that in the dynamical
system (\ref{GEND}), if the local map is continuous and bistable,
there are fronts with right continuous shape, fronts with left
continuous shape and unstable fronts, the various shapes only
differing at their discontinuity points.

Keeping on the assumption $c\in I_v^0$, 
one can modify the couplings in the following
way. We perturb ${\cal L}$ and ${\cal L'}$ with two
homogeneous operators in ${\cal M}$, namely $\tilde {\cal L}$ and  $\tilde
{\cal L'}$, of class $C^1$. Instead of (\ref{GEND}) we now consider
the following mapping on ${\cal M}$
\begin{equation}\label{NONLI}
u^{t+1}={\cal F}(u^t,\epsilon),
\end{equation}
where 
$$
{\cal F}(u,\epsilon)=\left((1-\epsilon){\cal L}+\epsilon\tilde {\cal L}\right)u
+\left((1-\epsilon){\cal L'}+\epsilon\tilde {\cal L'}\right) F\circ u.
$$
A front of velocity $v$ in this dynamical system is a travelling
wave for which the shape obeys the equation
$$
{\cal G}(u,\epsilon)\equiv {\cal F}(u,\epsilon)-\sigma^vu=0
$$
with the condition $H_c\circ u=H_0$ (where the Heaviside function is
the right continuous one). Its existence can be proved using a
continuation of the shape $\phi$ at $\epsilon=0$ (similarly as for
the continuation of solutions from the uncoupled limit in \cite{MacKay}).
\begin{Pro}
Given ${\cal L},{\cal L}',\tilde {\cal L}$ and $\tilde {\cal L}'$,
$v\in D(\phi(0,.))$ and $c\in I^0_v$, it exists $\epsilon_0$ such that
for all $\epsilon\in [0,\epsilon_0)$, the system (\ref{NONLI}) has
fronts of velocity $v$; their shape being a continuous function of $\epsilon$.
\end{Pro}
{\sl Proof:} For $\epsilon=0$ let $\phi$ be given by 
Theorem \ref{MAINR}. By assumption on $v$ and $c$ we have
$$
\inf_x|\phi(x)-c|=\delta,
$$
where $\delta$ is given by 
$$
\delta=\min\left\{c-\phi(0-0,v),\phi(0,v)-c\right\}>0.
$$ 
Let $U$ be the neighbourhood of $\phi$ in ${\cal M}$ defined as the set
of functions $u$ such that 
$$
\exists x_0\in\Bbb R\ :\ \|\phi-\sigma^{x_0}u\|<\delta,
$$
hence for $u\in U$, we have
$$
\left\{\begin{array}{ccl}
u(x)< c&{\rm if}&x< x^0\\
u(x)> c&{\rm if}&x \geq x^0
\end{array}\right. .
$$
In $U$, ${\cal G}$ is $C^1$ since $\tilde {\cal L}$, $\tilde {\cal L}'$ and
$F$ are. Moreover $D_u{\cal G}(\phi,0)$ 
is invertible and its inverse is bounded as follows from solving the
front shape equation in (\ref{GEND}). By the implicit function
theorem it exists $\epsilon_0$ and a unique continuous function
$u(\epsilon)$ such that $u(0)=\phi$ and for any
$\epsilon\in [0,\epsilon_0)$, we have $u(\epsilon)\in U$ and ${\cal
G}(u(\epsilon),\epsilon)=0$. \hfill $\Box$

Finally we notice that no supplementary difficulties are found when one
combines the two previous extensions, that is to say 
when one both perturbs the local map and the couplings.
\bigskip

\noindent {\large {\bf Acknowledgments}}

\noindent B.F.\ thanks the Department of Mathematics of the IST and
the Mathematical Physics Group of the Lisbon University for their
hospitality and their financial support during the period this work was
partly accomplished. He is supported by an EC grant on ``Nonlinear
dynamics and statistical physics of extended systems''.

\vfill\eject

\appendix

\section{Proof of Lemma \ref{POSS}}\label{FPROO}

\noindent
In all the proof, we will employ the sequences
$\{k_n\}_{n\in\Bbb N}$, $k_n\in \Bbb N$ and
$\{\epsilon_n\}_{n\in\Bbb N}$, $\epsilon_n \in \Bbb R$, $\epsilon_n>0$ 
so that 
$$
\lim_{n\rightarrow +\infty}k_n=+\infty \quad {\rm and}\quad 
\lim_{n\rightarrow +\infty}\epsilon_n=0.
$$

\noindent
{\em (i)} By assumption let $a,b\in E(h_1)$ be such that $a<b$ and
let $d\in E(h_2)$. 
Since $\theta$ is irrational 
the sequences $k_n$ and $\epsilon_n$ can be chosen
so that
$$
-\epsilon_n< -\left\{(k_n+1)\theta -\frac{d-a}{b-a}\right\}\leq 
0\quad \forall n\in \Bbb N,
$$
where $\{.\}$ stands for the fractional part of a real number \cite{Graham}.
Moreover since $0<\theta <1$ the sequence
$$
j_n=\lfloor (k_n+1)\theta -\frac{d-a}{b-a}\rfloor ,
$$
satisfies the inequalities 
$0\leq j_n\leq k_n$ for $n$ sufficiently large, say $n \geq
n_0$. Here $\lfloor x\rfloor$ denotes for the floor function
\cite{Graham}. Using the equality $x=\lfloor x\rfloor +\{x\}$, we obtain 
$$
-(b-a)\epsilon_n< d+(k_n-j_n)a+j_n b-(k_n+1)\left(a+\theta
(b-a)\right)\leq 0\quad \forall n\geq n_0,
$$
that is to say $a+\theta (b-a)\in P$.

\noindent
{\em (ii)} From relation (\ref{INCLU}) let us assume that 
$$v\in P\setminus \bigcup_{k=0}^{\infty}\frac{E(h_2)++_kE(h_1)}{k+1}.
$$
Then there exist the sequences 
$\{a_{i,n}\}_{n\in \Bbb N, 1\leq i\leq k_n}$, $a_{i,n}\in
E(h_1)$, and $\{d_n\}_{n\in\Bbb N}$, $d_n\in E(h_2)$ such that 
$$
-\epsilon_n< \frac{1}{k_n+1}\left(d_n+\sum_{i=1}^{k_n}a_{i,n}\right)
-v\leq 0\quad \forall n\in \Bbb N .
$$
Using the definition of $i_j$ and $s_j$, we have
$$
\frac{i_2+k_n i_1}{k_n+1}\leq
\frac{1}{k_n+1}\left(d_n+\sum_{i=1}^{k_n}a_{i,n}\right) 
\leq \frac{s_2+k_n s_1}{k_n+1}\quad \forall n\in \Bbb N.
$$
Consequently if $E(h_1)$ and $E(h_2)$ are bounded then $i_1\leq v\leq s_1$.
In this case one has
$$
E(\phi(0,.))\cap
[i_1,s_1]^c=\bigcup_{k=0}^{\infty}\frac{E(h_2)++_kE(h_1)}{k+1}\cap
[i_1,s_1]^c,
$$
and {\em (ii)} is proved.

\noindent
{\em (iii)} For $\theta >1$, $\theta$ irrational and $v=a+\theta (b-a)$
where $a,b\in E(h_1)$ are such that $a<b$, let $\{d_n\}_{n\in \Bbb N}$
be an increasing sequence of elements of $E(h_2)$ such that 
$$\lim_{n\rightarrow +\infty}d_n=+\infty.$$
Let us define 
$$
\beta_n=\frac{d_n-v}{b-a}\quad {\rm and}\quad 
\tilde{k}_n=\lceil \frac{\beta_n}{\theta}\rceil,
$$
where $\lceil x\rceil$ stands for the ceiling function \cite{Graham}.
By compactness it exists $\alpha\in [0,\theta]$ and a strictly increasing 
sequence of integers $\{n_i\}_{i\in \Bbb N}$ such that 
$$
\alpha =\lim_{i\rightarrow +\infty}\tilde{k}_{n_i}\theta-\beta_{n_i}.
$$
Let also $\{m_j\}_{j\in \Bbb N}$, $m_j\in \Bbb N$ be such that 
$$
\frac{\epsilon_j}{2}\leq \{m_j\theta +\alpha\}< \frac{2\epsilon_j}{3}.$$
The increase of $\beta_n$ results in the following. For all $j$
it exists $n_{i_j}$ such that 
$$
-1+\frac{\beta_{n_{i_j}}+1}{\theta-1}-\frac{\beta_{n_{i_j}}}{\theta}\geq
m_j \quad {\rm and}\quad 
|\tilde{k}_{n_{i_j}}\theta-\beta_{n_{i_j}}-\alpha |<\frac{\epsilon_j}{3}.
$$
Finally let $k_j=\tilde{k}_{n_{i_j}}+m_j$. 
From the previous inequalities, we obtain the following 
$$
0\leq \lfloor k_j\theta -\beta_{n_{i_j}}\rfloor \leq k_j
$$
and
$$
0\leq \{k_j\theta-\beta_{n_{i_j}}\}< \epsilon_j.
$$
Now let $p_j=\lfloor k_j\theta -\beta _{n_{i_j}}\rfloor $
to obtain 
$$
-(b-a)\epsilon_j<d_{n_{i_j}}+(k_j-p_j)a+p_jb-(k_j+1)
\left(a+\theta (b-a)\right) \leq 0,
$$
and then to conclude that $v\in P$.

\section{Proof of Lemma \ref{LDEN}}\label{TPROO}
Lemma \ref{LDEN} is a consequence of the combination of the following
statements.
\begin{Lem}\label{AUX1}
(i) If $D(h_2)\neq\emptyset$ and $E(h_1)=D(h_1)$ then $D(\phi(0,.))$ 
is dense in $(i_1,s_1)$.

\noindent
(ii) If $\sup D(h_2)=+\infty$ and $a\in D(h_1)$ then $D(\phi(0,.))$ is dense
in $(a,+\infty )$.

\noindent
(iii) If $\inf D(h_2)=-\infty$ and $a\in D(h_1)$ then $D(\phi(0,.))$ is dense
in $(-\infty ,a)$.
\end{Lem}
\begin{Lem}\label{AUX2}
If $D(\phi (0,.))$ is dense in $(v_1,v_2)$ then $G$ is nowhere
dense in $\bar{v}^{-1}\left((v_1,v_2)\right) $.
\end{Lem}

\noindent
{\sl Proof of Lemma \ref{AUX1}:} Let $a,b\in D(h_1)$ and $d\in
D(h_2)$. We have 
$$
\forall 0\leq j\leq k\quad \frac{d+(k-j)a+jb}{k+1}\in D(\phi(0,.)).
$$
Hence for any $\theta\in [0,1]$ by choosing $j_k=\lfloor
k\theta\rfloor$ we obtain
$$
a+\theta (b-a)=\lim_{k\rightarrow\infty}\frac{d+(k-j_k)a+j_kb}{k+1},
$$
which proves {\em (i)}.

\noindent
Now let $\{d_n\}_{n\in\Bbb N}$, $d_n\in D(h_2)$ and 
${\displaystyle\lim_{n\rightarrow\infty}}d_{n}=+\infty$. For
$\theta>0$ by choosing $k_n=\lfloor\frac{d_n}{\theta}\rfloor$ we have
$$
\frac{d_n+k_na}{k_n+1}\in D(\phi(0,.)),
$$
and thus 
$$
a+\theta =\lim_{k\rightarrow\infty}\frac{d_n+k_na}{k_n+1},
$$
{\em (ii)} is proved and one can proceed similarly for {\em
(iii)}. \hfill $\Box $

To prove Lemma \ref{AUX2}, we need the following auxiliary lemma.
\begin{Lem}\label{AUX3}
If $\bar{v}(c_1)=\bar{v}(c_2)$ then $\left(c_1,c_2\right)\cap G=\emptyset$.
\end{Lem}
{\sl Proof:} Let $c_1<c_2$ be such that $\bar{v}(c_1)=\bar{v}(c_2)$. For
any $c\in (c_1,c_2)$ we obtain, by using the definition of the function
$\bar{v}$, $\bar{v}(c)=\bar{v}(c_1)=\bar{v}(c_2)$ and 
$$
\phi(0,\bar{v}(c))\geq c_2>c>c_1\geq\phi(0,\bar{v}(c)-0)=\phi(0-0,\bar{v}(c)),
$$
and hence $c\notin G$.\hfill $\Box$

\noindent
Now let $c_1<c_2$ be such that $c_1,c_2\in
G\cap\bar{v}^{-1}((v_1,v_2))$. Then either $\bar{v}(c_1)=\bar{v}(c_2)$
and by Lemma \ref{AUX3} $(c_1,c_2)\cap G=\emptyset$, or 
$\bar{v}(c_1)<\bar{v}(c_2)$. In the latter case by density of
$D(\phi(0,.))$ it exists $v^{*}\in (\bar{v}(c_1),\bar{v}(c_2))\cap
D(\phi(0,.))$. Let us choose $c_1^{*},c_2^{*}\in
(\phi(0,v^{*}-0),\phi(0,v^{*}))$ such that $c_1^{*}<c_2^{*}$. We have 
$(c_1^{*},c_2^{*})\subset (c_1,c_2)$ and
$\bar{v}(c_1^{*})=\bar{v}(c_2^{*})$. So again Lemma \ref{AUX3} implies
that $(c_1^{*},c_2^{*})\cap G=\emptyset$. In both cases $G$ is not
dense in $(c_1,c_2)$. \hfill $\Box $


\begin{thebibliography}{99}
\bibitem{Afrai1}  {V.S.} {Afraimovich} and {L.A.} {Bunimovich}, {\em
Density of defects and spatial entropy in extended systems}, Physica D
{\bf 80} (1995) 277--288.

\bibitem{Afrai2}  {V.S.} {Afraimovich} and {V.I.} {Nekorkin}, {\em Chaos of
travelling waves in a discrete chain of diffusively coupled maps}, Int.\ J.\
Bif.\ Chaos {\bf 4} (1994) 631--637.

\bibitem{Bates} {P.W.} {Bates}, {P.C.} {Fife}, {X.} {Ren} and {X.}
{Wang}, {\em Traveling waves in a convolution model of phase
transitions}, Arch.\ Rational Mech.\ Anal.\ {\bf 138} (1997) 105--136.

\bibitem{Carretero} {R.} {Carretero-Gonz\'alez}, {D.K.} {Arrowsmith}
and {F.} {Vivaldi}, {\em Mode-locking in CML}, Physica D {\bf 103}
(1997) 381--403.

\bibitem{Coutinho96} {R.} {Coutinho} and {B.} {Fernandez}, {\em Extended
symbolic dynamics in bistable CML: Existence and stability of fronts},
Physica D, {\bf 108} (1997) 60--80.

\bibitem{Coutinho2} {R.} {Coutinho} and {B.} {Fernandez}, {\em On the
global orbits in a bistable CML}, Chaos {\bf 7} (1997) 301--310. 

\bibitem{Graham} {R.L.} {Graham}, {D.E.} {Knuth} and {O.} {Patashnik},
{\em Concrete Mathematics}, Addison-Wiley (1989). 

\bibitem{Chaos} {K.} {Kaneko}, ed., Focus issue on Coupled Map
Lattices, Chaos {\bf 2} 3 (1992).

\bibitem{Kaneko} {K.} {Kaneko}, ed., {\em Theory
and applications of Coupled Map Lattices}, Addison-Wiley (1993).

\bibitem{Kolmo} {A.N.} {Kolmogorov} and {S.V.} {Fomin}, {\em Elements
of the theory of functions and of functional analysis}, Mir (1976).

\bibitem{MacKay} {R.S.} {MacKay} et {J.-A.} {Sepulchre}, {\em
Multistability in networks of weakly coupled bistable units}, Physica
D {\bf 82} (1995) 243--254. 

\bibitem{Oppo} {G.L.} {Oppo} and {R.} {Kapral}, {\em Domain growth and
nucleation in a discrete bistable system}, Phys.\ Rev. A {\bf 36}
(1987) 5820--5831.

\bibitem{Poussin} {C.} de la Vall\'ee Poussin, {\em Int\'egrales de Lebesgue.
Fonctions d'Ensemble. Classes de Baire}, Gauthier-Villars et C$^{\rm ie}$
(1916).
\end{thebibliography}
\end{document}